\documentclass[11pt]{article}

\usepackage{amssymb}
\usepackage{amsmath}
\usepackage{epsfig}
\usepackage{cite}
\usepackage{verbatim}
\usepackage{enumerate}
\usepackage{subfigure}
\usepackage{multicol}
\usepackage{picinpar}
\usepackage{pstricks}
\usepackage{pst-node}

\def\QED{\mbox{\rule[0pt]{1.5ex}{1.5ex}}}

\newtheorem{theorem}{\bf Theorem}

\newtheorem{lemma}{\bf Lemma}

\begin{document}
\setcounter{page}{1}
\title{{Duality and Capacity Region of AF Relay MAC and BC}}
\author{\normalsize Syed A. Jafar and Krishna S. Gomadam\\
\small Center for Pervasive Communications and Computing\\
\small Department of Electrical Engineering and Computer Science\\ 
\small University of California Irvine, CA 92697 USA\\
{\small E-mail~:~ \{syed, kgomadam\}@uci.edu}}
\date{}
\maketitle
\thispagestyle{empty}
\begin{abstract}
We consider multi-hop multiple access (MAC) and broadcast channels (BC) where communication takes place with the assistance of  relays that  amplify and forward (AF) their received signals. For a two hop parallel AF relay MAC, assuming a sum power constraint across all relays we characterize optimal relay amplification factors and the resulting capacity regions. We find the sum rate capacity and the maximum rate for each user in closed form and express the optimal rate pair $(R_1,R_2)$ that maximizes $\mu_1R_1+\mu_2R_2$ as the solution of a pair of simultaneous equations. We find that the parallel AF relay MAC with total transmit power of the two users $P_1+P_2=P$ and total relay power $P_R$ is the dual of the parallel AF relay BC where the MAC source nodes become the BC destination nodes, the MAC destination node becomes the BC source node, the dual BC source transmit power is $P_R$ and the total transmit power of the AF relays is $P$. The duality means that the capacity region of the  AF relay MAC with a sum power constraint $P$ on the transmitters is the same as that of the dual BC.  The duality relationship is found to be useful in characterizing the capacity region of the AF relay BC as the union of MAC capacity regions. The duality extends to distributed relays with multiple antennas and more than $2$ hops as well. For the $M$ hop AF relay MAC where the $m^{th}$ hop consists of $R_m$ relays (possibly with multiple antennas) with total transmit power $P_m$ and source power is $P_0$, the dual BC is obtained by reversing the direction of communication so that the transmit power of the BC source is $P_M$ and the relays participating in the $i^{th}$ hop have transmit power $P_{M-i}$. Thus, even though the transmitters and receivers are switched, each link in the original MAC is associated with the same transmit power as in the dual BC.
\end{abstract}
\newpage
\section{Introduction}
Duality relationships are an important part of network information theory for the insights that they provide and the novel techniques that they inspire. In particular, wireless networks have the potential for many elegant duality relationships due to the inherent topological similarities between various cooperative communication scenarios. A case in point is the duality between the multiple input multiple output (MIMO) MAC and the MIMO BC \cite{Goldsmith_Jafar_Jindal_Vishwanath,DimacsViswanath}. It is shown that a MIMO MAC with a \emph{sum} power constraint $P$ across all transmitters has the same capacity region as its reciprocal network, i.e. the network obtained by switching the transmitters and receivers. The reciprocal network for the MIMO MAC is the MIMO BC with the same channels and transmit power $P$ available to the BC transmitter. Thus the MAC-BC duality relationship unifies the capacity results of the multiple access and broadcast channels. The MAC-BC duality is also instrumental in proving the optimality of dirty paper coding on the MIMO BC with Gaussian codebooks \cite{Dimacs,DimacsViswanath}, and has played a key role in one of the rare characterizations of the capacity region of a non-degraded broadcast channel \cite{Steinberg_Weingarten_Shamai, Mohseni_Cioffi}. Moreover, from a computational perspective this duality relationship enables the transformation of non-convex rate optimization problems on the broadcast channel into convex rate optimization problems on the dual multiple access channel \cite{Jindal_Jafar_Vishwanath_Goldsmith}. These simplifications lead to efficient resource allocation algorithms on the broadcast channel. 

The MAC-BC duality is especially interesting as it hints at a fundamental reciprocity property in a multiuser framework. The reciprocity of a wireless channel is a rudimentary property for point to point (single user) communications: i.e., the channel capacity is unchanged if the roles of the transmitter and receiver are switched. With multiple nodes the implications of switching the roles of transmitters and receivers are not easily characterized. The MAC-BC duality shows that for the single hop Gaussian MIMO MAC and BC such a reciprocity holds. It also confirms the intuition that the topological isomorphism between reciprocal wireless channels should translate into simple relationships between their capacity regions.


The MIMO MAC and BC model the uplink and the downlink for one-hop communication over cellular networks or a similar Wi-Fi network topology. While one-hop communication continues to be an active area of research, there is also increasing  interest in communication over multiple hops, i.e. relay networks, spurred by the apparent remarkable potential of cooperative communications. For example,  using fixed relays in cellular systems can significantly improve the communication capabilities of the network \cite{Tang_Chae_Heath_Cho}. Various relay strategies have been studied in literature, primarily for point to point communications. These strategies include amplify-and-forward \cite{laneman_worell_exploit,laneman_mod_det}, where the relay sends a scaled version of its received signal to the destination; demodulate-and-forward (DemF)\cite{laneman_mod_det}, where the relay demodulates individual symbols and retransmits those symbols without regard to an outer code; decode-and-forward (DecF) \cite{laneman_efficientprotocols}  where the relay decodes the entire message, re-encodes it and sends it to the destination; compress-and-forward (CF) \cite{cap_theorems_cover, cap_theorems_relay_kramer} where the relay sends a quantized version of its received signal; and estimate and forward (EF) where the relay sends a soft estimate of its received symbol to the destination. Of these, amplify and forward (AF) is perhaps the most interesting from a practical standpoint as it requires the relays to only scale their received symbols. Thus the complexity and cost of relaying, always an issue in designing cooperative networks, is minimal for AF relay networks. In addition to its simplicity AF is known to be the optimal relay strategy in many interesting cases\cite{cap_lar_gauss_relay, gastpar_infocom, elgamaal_linear, break_net,relay_without_delay}. It is also shown that the power efficiency of AF relaying increases with the number of relays in a random network \cite{Dana_Hassibi}.

AF relay optimization for dual hop communications has been the focus of much recent research. \cite{Serbetli_Yener, Chen_Serbetli_Yener,Lee_Yener} consider the case of orthogonal relay transmissions. While orthogonal relay schemes are attractive for wideband communications, Maric and Yates \cite{Maric_Yates} have shown that for amplify and forward relays, shared bandwidth transmission schemes can provide higher capacity. Maric and Yates also find closed form solutions for the relay amplification factor and the point to point AF relay channel capacity with shared band transmission. Optimum and near optimum power allocation schemes for single branch multi-hop relay networks have been considered in \cite{Hasna_Alouini}.

While much of the work on relay networks has focused on point to point communications, multiuser relay networks are increasingly gaining attention as seminal work in this area \cite{Gupta_Kumar, cap_theorems_relay_kramer, Tang_Chae_Heath_Cho,distributed_mmse_nima_sayed,Liang_Veeravalli} has shown the remarkable advantages of multiuser relaying. In \cite{distributed_multiuser_mmse_berger_wittenben} gains for AF relays in a multiuser parallel network are determined to achieve a joint minimization of the MMSE of all the source signals at the destination. Tang et. al. \cite{Tang_Chae_Heath_Cho} consider a MIMO relay broadcast channel, where a multiple antenna transmitter sends data to multiple users via a relay with multiple antennas over two hops. They find different algorithms for computing the transmit precoder, relay linear processing matrix and the sum rate under the assumption of zero-forcing dirty paper coding and Gaussian signals. Capacity bounds are used to establish that the performance loss is not significant. Capacity with cooperative relays has been explored for the multicast problem by Maric and Yates \cite{Maric_Yates_CB,Maric_Yates_MB}, for the broadcast problem by  Liang and Veeravalli \cite{Liang_Veeravalli}, and for the mixed multiple access and broadcast problem by Host-Madsen \cite{MadsenIT}. Maric and Yates explore an accumulative multicast strategy where nodes collect energy from previous transmissions, while Liang and Veeravalli \cite{Liang_Veeravalli} and Host-Madsen \cite{MadsenIT} address the general question of optimal relay functionality which may not be an amplify and forward scheme.

In this work, we pursue two related objectives. The first is to investigate the capacity optimal relay amplification factors for two hop multiple access and broadcast channels where communication takes place via parallel AF relay links and no direct link exists between source(s) and destination(s). With a sum power constraint on all the relays, we characterize the optimal relay amplification factor for the parallel AF relay MAC for all points on the boundary of the MAC capacity region. We obtain the sum capacity and the individual user capacities in closed form and present a couple of simultaneous equations whose numerical solution yields the optimal rate pair $(R_1,R_2)$ that maximizes $\mu_1R_1+\mu_2R_2$ for any $\mu_1,\mu_2\geq 0$. The capacity region of the parallel AF relay BC optimized over all relay amplification factors is found to be non-convex and is harder to obtain directly. However, we identify an interesting duality relationship that allows us to compute the capacity region of the parallel AF relay BC as the union of the relay MAC capacity regions.

The second objective of this work is to identify duality relationships in AF relay networks. We  obtain a general duality result for multi-hop multiple access and broadcast channels where each hop may consist of parallel AF relays and the relays may be equipped with multiple antennas. For the two hop case, we show that the multiple access channel with total transmit power of all users equal to $P$ and total relay transmit power $P_R$ is the dual of the BC obtained when the destination becomes the transmitter and the transmitters become the receivers, and the powers are switched as well, i.e. in the dual BC, the transmit power is $P_R$ and the total relay power is $P$. In general, for multi hop AF relay networks with parallel AF relays at each hop, and with possibly multiple antennas at the relays, the duality result holds when each link is associated with the same transmit power in the original MAC and the dual BC channels.

\section{Single User AF Parallel Relay Channel}\label{sec:ptp}
We start with the simple illustrative example of a single user (point to point) parallel AF relay channel.
\subsection{Point to Point Channel Model: PTP$({\bf F},P,{\bf G}, \{{\bf D}\},P_R)$ }
The input-output equations for this case are as follows:
\begin{eqnarray}
{\bf R} &=& {\bf F}x+{\bf N}_R\\
y&=& \mbox{Tr}[{\bf G} {\bf D}\left({\bf F}x+{\bf N}_R \right)]+n
\end{eqnarray}
where ${\bf R}, {\bf F}, {\bf G}, {\bf D}$ and ${\bf N}_R$ are $R\times R$ \emph{diagonal} matrices with the $i^{th}$ principal diagonal terms $r_i,~ f_i,~ g_i,~ d_i,~ n_{i,R}$ respectively representing the received signal at the  $i^{th}$ relay node, the channel coefficient from the source to the $i^{th}$ relay node, the channel coefficient from the $i^{th}$ relay node to the destination, the relay amplification factor for the $i^{th}$ relay node, and the additive white Gaussian noise (AWGN) component at the $i^{th}$ relay receiver, modeled as an i.i.d. zero mean unit variance Gaussian random variable. $x$ is the source transmitted symbol and $n$ is the zero mean unit variance AWGN at the destination node. The power constraints are specified as follows:
\begin{eqnarray}
\mbox{Source Power Constraint:}&~~~~&\mbox{E}[x^2]=P,\\
\mbox{Relay Power Constraint:}&~~~~&\mbox{E}[|| {\bf D}\left({\bf F}x+{\bf N_R} \right) ||^2]=\mbox{Tr}({\bf D}^2(I+P{\bf F}^2))=P_R.
\end{eqnarray}

The point to point channel under these assumptions is denoted as PTP$({\bf F},P,{\bf G}, \{{\bf D}\},P_R)$. We use the notation $\{{\bf D}\}$ to indicate all feasible relay amplification matrices while we use ${\bf D}$ to indicate a specific choice of the relay amplification matrix. For example, PTP$({\bf F},P,{\bf G}, {\bf D},P_R)$ refers to the point to point channel with a given ${\bf D}$ matrix.

We assume fixed channels so that all the channel knowledge is available to all nodes. Note that while in general individual power constraints at relay nodes may be a more reasonable assumption for distributed nodes, we assume a sum power constraint across all relay nodes instead. This assumption may be more relevant for fixed wireless networks with slowly varying channels where the channel information overhead is small. The sum power constraint is interesting for two reaons. First, it provides useful insights into relay optmization. Optimal resource allocation, e.g., optimizing relay amplification factors is a challenging problem especially when the relays serve multiple users simultaneously, as in multiple access or broadcast channels. The second reason is that the sum power constraint enables powerful duality relationships that are also of fundamental interest.

Finally, throughout this paper we consider real signals, noise, channel and relay amplification factors. However, most of the results apply to the complex case as well. For example, for the point to point channel, it is easily seen that even with complex quantities, the relay must choose the phase of its amplification factor $d_i$ to cancel the total phase of $f_i$ and $g_i$. Therefore, there is no loss of generality in the assumption that $f_i, g_i, d_i$ are real. The exceptions will be pointed out in the conclusions section. 

\subsection{Capacity and Optimal Relay Amplification}
The capacity of the point to point parallel AF relay channel and the optimal relay amplification factors have been obtained previously by Maric and Yates \cite{Maric_Yates}. We re-visit the result in order to introduce our notation and to illustrate the key ideas that are later applied to multiple access channels. Given any choice of relay amplification vector ${\bf D}$ that satisfies the relay transmit power constraint, the resulting channel is an AWGN channel whose capacity is simply  
\begin{eqnarray}
C^{PTP}({\bf F},P,{\bf G}, {\bf D},P_R)&=&\log\left(1+\frac{\left[\mbox{Tr}\left({\bf G}{\bf D}{\bf F}\right)\right]^2P}{1+\mbox{Tr}({\bf D}^2{\bf G}^2)}\right). 
\end{eqnarray}
Therefore, the capacity optimization problem for the point to point channel may be represented as:
\begin{eqnarray}
C^{PTP}({\bf F},P,{\bf G}, \{{\bf D}\},P_R)&=&\max_{{\bf D}\in\mathcal{D}} C^{PTP}({\bf F},P,{\bf G}, {\bf D},P_R)
\end{eqnarray}
where $\mathcal{D}$ is the set of feasible relay amplification matrices:
\begin{eqnarray*}\label{eq:ptppower}
\mathcal{D}&=&\left\{{\bf D}: \mbox{Tr}({\bf D}^2(I+P{\bf F}^2))=P_R\right\}
\end{eqnarray*}

The following theorem presents the closed form capacity expression.
\begin{theorem}\label{theorem:ptpcapacity} {\bf [Maric, Yates 2004]}
The capacity of the point to point channel described above is 
\begin{eqnarray*}
C^{PTP}(P,{\bf F},\{{\bf D}\},P_R, {\bf G}) &=& \log\left(1+PP_R \mbox{Tr}\left[{\bf G}^2{\bf F}^{2}\left(I+P{\bf F}^{2}+P_R{\bf G}^2\right)^{-1}\right]\right)\\
&=& \log\left(1+PP_R\sum_{k=1}^R\frac{f_k^{2}g_k^2}{1+Pf_k^{2}+ P_R g_k^2}\right).
\end{eqnarray*}
The optimum relay amplification vector for the point to point relay channel is given by
\begin{eqnarray}
{\bf D} = \gamma {\bf F}{\bf G}\left(I+P {\bf F}^{[1]2}+P_R {\bf G}^2\right)^{-1} \label{eq:ptpD}
\end{eqnarray}
where $\gamma$ is a constant necessary to satisfy the relay transmit power constraint, and may be expressed explicitly as:
\begin{eqnarray}
\gamma  =  \sqrt{\frac{P_R}{\mbox{Tr}\left({\bf F^{2}}{\bf G}^2\left(I+P {\bf F}^{2}+P_R {\bf G}^2\right)^{-2}\left(I+P {\bf F}^{2}\right)\right)}}
\end{eqnarray}
\end{theorem}
The proof of this result is provided by Maric and Yates in \cite{Maric_Yates}.

\subsection{Reciprocity of the Point to Point AF Relay Channel}

\begin{figure}[h]
\input{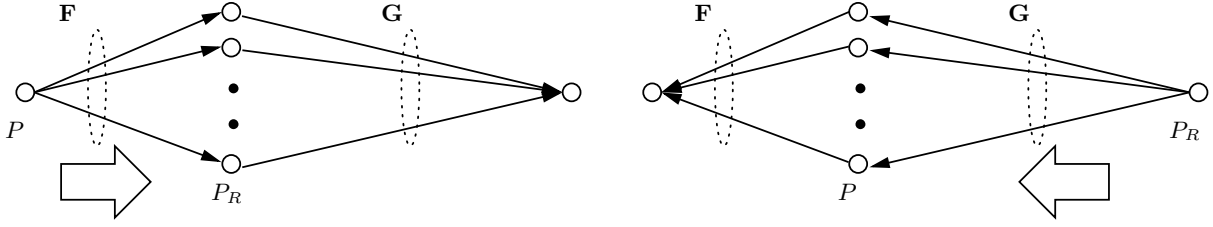}
\caption{Point to point dual channels}\label{fig:ptpdual}
\end{figure}

From the result of Theorem \ref{theorem:ptpcapacity}, notice that
\begin{eqnarray}
C^{PTP}({\bf F},P,{\bf G},\{{\bf D}\},P_R) &=& C^{PTP}({\bf G},P_R,{\bf F},\{{\bf D}\},P) 
\end{eqnarray}
The capacity is unchanged if we switch variables as follows:
\begin{eqnarray}
P&\rightarrow& P_R\\
P_R&\rightarrow& P\\
{\bf F}&\rightarrow& {\bf G}\\
{\bf G}&\rightarrow& {\bf F}
\end{eqnarray}
In other words, the capacity is the same if we switch the transmitter and the receiver as long as the powers of the transmitter and the relay are also switched. Fig. \ref{fig:ptpdual} shows the dual channels that have the same capacity.

Further, note that the optimal relay amplification vector was found as
\begin{eqnarray}
{\bf D} = \gamma {\bf F^{}{\bf G}\left(I+P {\bf F}^{2}+P_R {\bf G}^2\right)^{-1}} 
\end{eqnarray}
Note that except for the constant $\gamma$ this matrix is also unchanged when applied to either the original channel or the dual channel. However, $\gamma$ must be different on the two channels because the power constraints are different. For the original channel $\gamma$ is given by Theorem \ref{theorem:ptpcapacity}, whereas on the dual channel we will have
\begin{eqnarray}
\gamma_{\mbox{dual}}  =  \sqrt{\frac{P}{\mbox{Tr}\left({\bf F^{2}{\bf G}^2\left(I+P {\bf F}^{2}+P_R {\bf G}^2\right)^{-2}}\left(I+P {\bf G}^2\right)\right)}}
\end{eqnarray}
The following theorem shows that this duality property is actually much stronger as it holds not only for the optimal relay amplification matrix ${\bf D}$ but rather for \emph{every} feasible ${\bf D}$.
\begin{theorem}\label{theorem:ptpdual}
\begin{eqnarray}
C^{PTP}({\bf F},P,{\bf G},{\bf D},P_R) &=& C^{PTP}({\bf G},P_R,{\bf F},\kappa{\bf D},P) 
\end{eqnarray}
Given any relay amplification matrix ${\bf D}$ for a point to point parallel AF relay channel PTP$({\bf F},P,{\bf G},{\bf D},P_R)$ there exists a dual point to point parallel AF relay channel PTP$({\bf G},P_R,{\bf F},\kappa{\bf D},P) $ that has the same capacity as PTP$({\bf F},P,{\bf G},{\bf D},P_R)$ and where $\kappa$ is chosen to satisfy the relay power constraint of the dual channel.
\end{theorem}
\begin{proof}
We start with the capacity and relay power constraint expressions of the original and the dual point to point channels.
\begin{eqnarray}
&C^{PTP}({\bf F},P,{\bf G},{\bf D},P_R)=\log\left(1+\frac{\left[\mbox{Tr}\left({\bf G}{\bf D}{\bf F}\right)\right]^2P}{1+\mbox{Tr}({\bf D}^2{\bf G}^2)}\right), & ~~~~\mbox{Tr}({\bf D}^2(I+P{\bf F}^2))=P_R\\
&C^{PTP}({\bf G}, P_R, {\bf F}, \kappa{\bf D}, P)=\log\left(1+\frac{\left[\mbox{Tr}\left({\bf F}{\kappa\bf D}{\bf G}\right)\right]^2P_R}{1+\mbox{Tr}(\kappa^2{\bf D}^2{\bf F}^2)}\right), & ~~~~\mbox{Tr}(\kappa^2{\bf D}^2(I+P_R{\bf G}^2))=P
\end{eqnarray}
Substituting from the power constraint into the capacity expression we have:
{\allowdisplaybreaks
\begin{eqnarray}
C^{PTP}({\bf F},P,{\bf G},{\bf D},P_R)&=&\log\left(1+\frac{\left[\mbox{Tr}\left({\bf G}{\bf D}{\bf F}\right)\right]^2P}{\frac{\mbox{Tr}({\bf D}^2(I+P{\bf F}^2))}{P_R}+\mbox{Tr}({\bf D}^2{\bf G}^2)}\right)\\
&=& \log\left(1+\frac{\left[\mbox{Tr}\left({\bf G}{\bf D}{\bf F}\right)\right]^2PP_R}{\mbox{Tr}({\bf D}^2(I+P{\bf F}^2+P_R{\bf G}^2)}\right)\\
C^{PTP}({\bf G}, P_R, {\bf F}, \kappa{\bf D}, P)&=&\log\left(1+\frac{\left[\mbox{Tr}\left({\bf F}{\kappa\bf D}{\bf G}\right)\right]^2P_R}{\frac{\mbox{Tr}(\kappa^2{\bf D}^2(I+P_R{\bf G}^2))}{P}+\mbox{Tr}(\kappa^2{\bf D}^2{\bf F}^2)}\right)\\
&=& \log\left(1+\frac{\left[\mbox{Tr}\left({\bf G}{\bf D}{\bf F}\right)\right]^2PP_R}{\mbox{Tr}({\bf D}^2(I+P{\bf F}^2+P_R{\bf G}^2)}\right)
\end{eqnarray}
}
Thus, $C^{PTP}({\bf F},P,{\bf G},{\bf D},P_R)=C^{PTP}({\bf G}, P_R, {\bf F}, \kappa{\bf D}, P)$.\hfill\QED
\end{proof}

The proof of Theorem \ref{theorem:ptpdual} illustrates an important point. By normalizing the noise power to unity and also incorporating the relay power constraint into the normalized channel definition PTP$({\bf F},P,{\bf G},{\bf D},P_R)$ may be represented as:
\begin{eqnarray}
y&=&\frac{\mbox{Tr}\left({\bf G}{\bf D}{\bf F}\right)\sqrt{P_R}}{\sqrt{\mbox{Tr}({\bf D}^2(I+P{\bf F}^2+P_R{\bf G}^2))}}x+n.
\end{eqnarray}
Comparing this to the normalized representation of the dual PTP$({\bf G}, P_R, {\bf F}, \kappa{\bf D}, P)$,
\begin{eqnarray}
y&=&\frac{\mbox{Tr}\left({\bf G}{\bf D}{\bf F}\right)\sqrt{P_R}}{\sqrt{\mbox{Tr}({\bf D}^2(I+P{\bf F}^2+P_R{\bf G}^2))}}x+n.
\end{eqnarray}
notice that the normalized dual channel is identical to the original channel.

The appealing feature of this normalized form is that it is unaffected by any scaling of the relay amplification matrix ${\bf D}$ by any constant $\kappa$. While there is only a unique scaling factor for any diagonal matrix such that it will satisfy the relay transmit power constraint, that scaling factor does not affect the normalized channel as the power constraint is already accommodated into this form. Further, the reciprocity of the point to point channel is also evident from the normalized form.
\section{Parallel AF Relay MAC and BC}
\begin{figure}[h]
\input{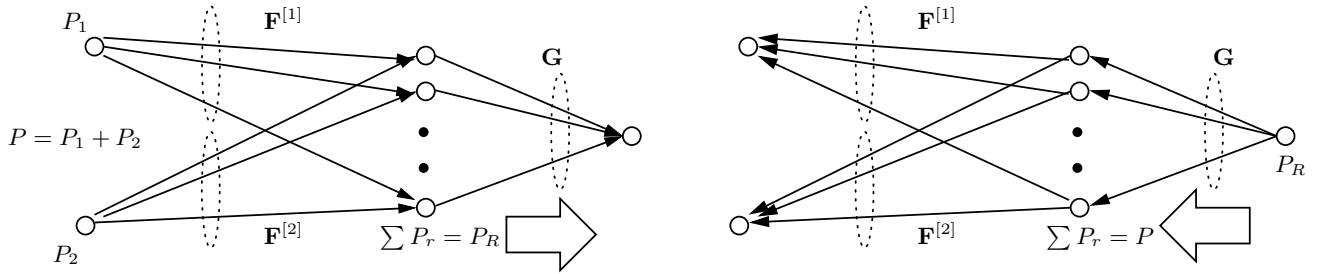}
\caption{Dual Parallel AF Relay MAC and BC channels}\label{fig:dualmacbc}
\end{figure}
We consider two hop multiple access and broadcast channels where communication takes place with the assistance of multiple parallel AF relays. As in the point to point case, all channels are known and fixed, there is one sum power constraint across all relays, and all channel, input, output, noise variables and the relay scaling factors are real. For simplicity we focus on the two user case. The AF relay MAC and its dual BC models are depicted in Fig. \ref{fig:dualmacbc}. 

\subsection{Symbol Definitions and Notation}
\begin{eqnarray*}
{\allowdisplaybreaks
\begin{array}{|c|c|c|}\hline
\mbox{Variable} & \mbox{AF MAC} & \mbox{Dual AF BC}\\\hline
R & \mbox{Number of parallel AF relays}&\mbox{Number of parallel AF relays}\\\hline
P_1 & \mbox{Transmit power for source $1$}&\\\hline
P_2 & \mbox{Transmit power for source $2$}&\\\hline
P & P_1+P_2 & \mbox{Total transmit power of $R$ relays}\\\hline
P_R & \mbox{Total transmit power of $R$ relays} & \mbox{Source transmit power}\\\hline
f^{[j]}_i & \mbox{Channel  from $j^{th}$ source to $i^{th}$ relay}&\mbox{Channel from $i^{th}$ relay to $j^{th}$ destination.} \\\hline
g_i & \mbox{Channel  from $i^{th}$ relay to destination}&\mbox{Channel from source to $i^{th}$ relay.}\\\hline
d_i & \mbox{Amplification factor at the $i^{th}$ relay}& \mbox{Amplification factor at the $i^{th}$ relay}\\\hline
x & &\mbox{Transmitted symbol for common source}\\\hline
x_i & \mbox{Transmitted symbol for source $i$}&\\\hline
y & \mbox{Received symbol at common destination} &\\\hline
y_i & &\mbox{Received symbol at $i^{th}$ destination}\\\hline
r_i & \mbox{Received symbol at the $i^{th}$ relay}& \mbox{Received symbol at the $i^{th}$ relay}\\\hline
d_ir_i & \mbox{Symbol transmitted by the $i^{th}$ relay}& \mbox{Symbol transmitted by the $i^{th}$ relay}\\\hline
n_{R,i} & \mbox{Unit power AWGN at the $i^{th}$ relay}& \mbox{Unit power AWGN at the $i^{th}$ relay}\\\hline
n_i & &\mbox{Unit power AWGN at $i^{th}$ destination}\\\hline
n & \mbox{Unit power AWGN at common destination}&\\\hline
\end{array}\nonumber\vspace{0.2cm}
}
\end{eqnarray*}

\subsection{Two  user Parallel AF Relay MAC and BC Models}
We use the notation MAC$({\bf F}^{[1]}, P_1, {\bf F}^{[2]}, P_2, {\bf G}, \{{\bf D}\}, P_R)$ to denote the two user multiple access channel described above, i.e. with the channels between the transmitters and relays ${\bf F}^{[1]}, {\bf F}^{[2]}$, the corresponding source transmit powers $P_1, P_2$ (respectively), the channel between the relays and the destination ${\bf G}$, the relay amplification vector ${\bf D}$, and total transmit power at all relays $P_R$. To distinguish the MAC resulting from a specific choice of ${\bf D}$ from the MAC where all {\bf D} from the feasible set are allowed, we denote the former as MAC$\left({\bf F}^{[1]}, P_1, {\bf F}^{[2]}, P_2, {\bf G}, {\bf D}, P_R\right)$ and the latter as MAC$\left({\bf F}^{[1]}, P_1, {\bf F}^{[2]}, P_2, {\bf G}, \left\{{\bf D}\right\}, P_R\right)$

\subsubsection{MAC$\left({\bf F}^{[1]}, P_1, {\bf F}^{[2]}, P_2, {\bf G}, \{{\bf D}\}, P_R\right)$}
For the two user parallel AF relay MAC$\left({\bf F}^{[1]}, P_1, {\bf F}^{[2]}, P_2, {\bf G}, \{{\bf D}\}, P_R\right)$, the received signals at the relays and the common destination are as follows:
\begin{eqnarray}
{\bf R} &=& {\bf F}^{[1]}x_1+{\bf F}^{[2]}x_2+{\bf N}_R\\
y&=& \mbox{Tr}[{\bf G} {\bf D}\left({\bf F}^{[1]}x_1+{\bf F}^{[2]}x_2+{\bf N}_R \right)]+n
\end{eqnarray}
Normalizing the noise to unit variance, the destination output can be expressed as
\begin{eqnarray}
y'&=& \frac{\mbox{Tr}\left({\bf G}{\bf D}{\bf F}^{[1]}\right)}{\sqrt{1+\mbox{Tr}\left({\bf D}^2{\bf G}^2\right)}}~~ x_1+\frac{\mbox{Tr}\left({\bf G}{\bf D}{\bf F}^{[2]}\right)}{\sqrt{1+\mbox{Tr}\left({\bf D}^2{\bf G}^2\right)}}~~ x_2+n'
\end{eqnarray}
The power constraints are:
\begin{eqnarray}\label{eq:powermac}
\mbox{E}[x_1^2]&=&P_1\\
\mbox{E}[x_2^2]&=&P_2\\
\mbox{Tr}\left({\bf D}^2\left(I+P_1{\bf F}^{[1]2}+P_2{\bf F}^{[2]2}\right)\right)& =&P_R.
\end{eqnarray}
As in the point to point case, the power constraint can be substituted into the channel output equation to obtain a corresponding normalized form 
\begin{eqnarray}
y'&=& \frac{\mbox{Tr}\left({\bf G}{\bf D}{\bf F}^{[1]}\right)\sqrt{P_R}}{\sqrt{\mbox{Tr}\left[{\bf D}^2\left(I+P_1{\bf F}^{[1]2}+P_2{\bf F}^{[2]2}+P_R{\bf G}^2\right)\right]}}~ x_1\nonumber\\
&&+\frac{\mbox{Tr}\left({\bf G}{\bf D}{\bf F}^{[2]}\right)\sqrt{P_R}}{\sqrt{\mbox{Tr}\left[{\bf D}^2\left(I+P_1{\bf F}^{[1]2}+P_2{\bf F}^{[2]2}+P_R{\bf G}^2\right)\right]}}~ x_2 + n'\label{eq:macnormal}
\end{eqnarray}

\subsubsection{BC$\left({\bf G}, P_R, {\bf F}^{[1]}, {\bf F}^{[2]}, \{{\bf D}\}, P\right)$}
We use the shorthand notation BC$\left({\bf G}, P_R, {\bf F}^{[1]}, {\bf F}^{[2]}, {\bf D}, P\right)$ to indicate the dual BC, i.e. broadcast channel with transmit power $P_R$, channel vector ${\bf G}$ from transmitter to relays, channel vectors ${\bf F}^{[1]}$ and ${\bf F}^{[2]}$ from the relays to receiver $1$ and $2$ respectively, relay amplification factor given by ${\bf D}$ and total transmit power used by the relays $P$. As for the MAC, we use $\{{\bf D}\}$ to indicate all feasible relay amplification factors are allowed and ${\bf D}$ to indicate a specific choice. For the dual broadcast channel, received signals at the relays and the two destinations are as follows:
\begin{eqnarray}
{\bf R} &=& {\bf G}x+{\bf N}_R\\
y_1&=& \mbox{Tr}[{\bf F}^{[1]} {\bf D}({\bf G}x+{\bf N}_R) ]+n_1\\
y_2&=& \mbox{Tr}[{\bf F}^{[2]} {\bf D}({\bf G}x+{\bf N}_R) ]+n_2
\end{eqnarray}
with power constraints
\begin{eqnarray}
\mbox{E}[x^2]&=&P_R\\
\mbox{Tr}\left({\bf D}^2\left(I+P{\bf G}^{2}\right)\right)& =&P.
\end{eqnarray}
Notice that the relays are associated with power $P$ and the source with power $P_R$.
Normalizing the AWGN to unit power,
{\allowdisplaybreaks
\begin{eqnarray}
y_1'&=& \frac{\mbox{Tr}\left({\bf F}^{[1]}{\bf D}{\bf G}\right)}{\sqrt{1+\mbox{Tr}\left({\bf D}^2{\bf F}^{[1]2}\right)}}~ x+n_1'\\
y_2'&=& \frac{\mbox{Tr}\left({\bf F}^{[2]}{\bf D}{\bf G}\right)}{\sqrt{1+\mbox{Tr}\left({\bf D}^2{\bf F}^{[2]2}\right)}}~ x+n_2'
\end{eqnarray}
}
Incorporating the relay power constraint into the normalized channel,
\begin{eqnarray}
y_1'&=& \frac{\mbox{Tr}\left({\bf F}^{[1]}{\bf D}{\bf G}\right)\sqrt{P}}{\sqrt{\mbox{Tr}\left[{\bf D}^2\left(I+P{\bf F}^{[1]2}+P_R{\bf G}^2\right)\right]}}~ x + n'\nonumber\\
y_2'&=& \frac{\mbox{Tr}\left({\bf F}^{[2]}{\bf D}{\bf G}\right)\sqrt{P}}{\sqrt{\mbox{Tr}\left[{\bf D}^2\left(I+P{\bf F}^{[2]2}+P_R{\bf G}^2\right)\right]}}~ x +n' \label{eq:bcnormal}
\end{eqnarray}
Recall that for the point to point case, the dual channel and the original channel were identical. Even for the conventional MAC-BC duality, the channels on the MAC and BC are identical. However, the forms of the equivalent normalized channels for the MAC and BC presented above are significantly different. In particular, while in the BC, user 1's channel is independent of ${\bf F}^{[2]}$ and user 2's channel independent of ${\bf F}^{[1]}$, in the MAC all channels depend on all parameters. A duality relationship between these two channels is therefore not a trivial observation based on their normalized channels. What makes the existence of a duality relationship even more unlikely is the fact that the channels themselves depend on how the total power is split among the users in the multiple access channel. With these apparent complexities, it is rather surprising that a precise duality relationship does exist between the MAC and BC described above, as we show in Section \ref{sec:macbcdual}.

\section{Capacity and Relay Optimization for Parallel AF Relay MAC}
Given a relay amplification vector ${\bf D}$ the capacity region of the resulting scalar Gaussian MAC is the well known pentagon that can be expressed as:
{\allowdisplaybreaks
\begin{eqnarray}
\mathcal{C}^{MAC}\left({\bf F}^{[1]}, P_1, {\bf F}^{[2]}, P_2, {\bf G}, {\bf D}, P_R\right) = \left\{(R_1,R_2)\right.&:&R_1 \leq \log\left(1+P_1\frac{\left[\mbox{Tr}\left({\bf D}{\bf G}{\bf F}^{[1]}\right)\right]^2}{1+\mbox{Tr}\left({\bf D}^2{\bf G}^2\right)}\right),\nonumber\\
R_2 &\leq& \log\left(1+P_2\frac{\left[\mbox{Tr}\left({\bf D}{\bf G}{\bf F}^{[2]}\right)\right]^2}{1+\mbox{Tr}\left({\bf D}^2{\bf G}^2\right)}\right),\nonumber\\
R_1+R_2&\leq&\left(1+\frac{P_1\left[\mbox{Tr}\left({\bf D}{\bf G}{\bf F}^{[1]}\right)\right]^2 +P_2\left[\mbox{Tr}\left({\bf D}{\bf G}{\bf F}^{[2]}\right)\right]^2}{1+\mbox{Tr}\left({\bf D}^2{\bf G}^2\right)}\right)\left.\right\}.\nonumber
\end{eqnarray}
}
Taking the union over all ${\bf D}$ that satisfy the relay sum power constraint gives us a characterization of the capacity region of MAC$\left({\bf F}^{[1]}, P_1, {\bf F}^{[2]}, P_2, {\bf G}, \{{\bf D}\}, P_R\right)$ as:
\begin{eqnarray}
\mathcal{C}^{MAC}\left({\bf F}^{[1]}, P_1, {\bf F}^{[2]}, P_2, {\bf G}, \{{\bf D}\}, P_R\right) &= & \cup_{{\bf D}\in\mathcal{D}_{MAC}}\mathcal{C}^{MAC}\left({\bf F}^{[1]}, P_1, {\bf F}^{[2]}, P_2, {\bf G}, {\bf D}, \phi\right)
\end{eqnarray}
where
\begin{eqnarray}
\mathcal{D}_{MAC}=\left\{{\bf D}: \mbox{Tr}\left({\bf D}^2\left(I+P_1{\bf F}^{[1]2}+P_2{\bf F}^{[2]2}\right)\right)=P_R\right\}
\end{eqnarray}
is the set of all relay amplification factors that satisfy the total transmit power constraint at the relays. Note that in the absence of any further characterization of ${\bf D}$ we are left with optimization over the entire space of feasible ${\bf D}$, i.e. a $R$ dimensional space. A brute force solution to such an optimization may be difficult as the number of relays increases. Theorem \ref{theorem:maccap} solves this problem.

\subsection{Relay Optimization}
The following Theorem reveals the structure of the optimal relay amplification factor ${\bf D}$ for rate pairs on the boundary of the capacity region.
\begin{theorem}\label{theorem:maccap}
The optimal relay amplification matrix ${\bf D}$ to maximize any weighted sum of users' rates $\mu_1R_2+\mu_2R_2$ with $\mu_1,\mu_2\geq 0$, has the following form:
\begin{eqnarray}
{\bf D}(\theta) = \gamma {\bf G}\left(P_1{\bf  F}^{[1]}\sin\theta+P_2 {\bf F}^{[2]}\cos\theta\right)\left(I+P_1 {\bf F}^{[1]2}+P_2 {\bf F}^{[2]2}+P_R {\bf G}^2\right)^{-1}
\end{eqnarray}
and $\gamma$  may be expressed explicitly as:
\begin{eqnarray}
\gamma  =  \sqrt{\frac{P_R}{\mbox{Tr}\left({\bf G}^2\left(P_1{\bf  F}^{[1]}\sin\theta+P_2 {\bf F}^{[2]}\cos\theta\right)^2\left(I+P_1 {\bf F}^{[1]2}+P_2 {\bf F}^{[2]2}+P_R {\bf G}^2\right)^{-2}\left(I+P_1 {\bf F}^{[1]2}+P_2 {\bf F}^{[2]2}\right)\right)}}\nonumber
\end{eqnarray}
\end{theorem}
The proof of Theorem \ref{theorem:maccap} is presented in Appendix \ref{proof:maccap}. Note that the optimization space over ${\bf D}$ is now only one dimensional, as opposed to the original $R$ dimensional space. To identify a point on the boundary of the capacity region one only needs the corresponding $\theta$. Therefore, the angle $\theta$ in Theorem \ref{theorem:maccap} plays a very important role. As we will establish in the following Theorems, $|\theta|$ going from $0$ to $\pi/2$ describes the boundary of the capacity region, with $\theta=0$ corresponding to the point where user $2$ achieves his maximum rate, $\theta^{11}$ (defined in Theorem \ref{theorem:sumrate}) corresponds to the points where the sum rate is maximized, and $|\theta|=\pi/2$ corresponds to the point where user $1$ achieves his maximum rate. This is also depicted in Fig \ref{fig:theta}.

Next we characterize the capacity region explicitly by obtaining in closed form the rate pairs corresponding to the sum capacity as well as the individual user capacities. We will also present a system of equations whose solution is the rate pair $(R_1,R_2)$ that maximizes $\mu_1 R_1+\mu_2 R_2$ for any positive $\mu_1+\mu_2$. We start with the extreme point $\mu_1=1, \mu_2=0$ that characterizes the maximum possible rate of user $1$.

\subsection{Maximum Individual Rate $R_1$}
\begin{theorem}\label{theorem:mu1}
For the MAC$\left({\bf F}^{[1]}, P_1, {\bf F}^{[2]}, P_2, {\bf G}, \{{\bf D}\}, P_R\right)$, the maximum rate $R_1$ that can be supported is
{\allowdisplaybreaks
\begin{eqnarray}
C_1^{10}&=& \log\left(1+P_1P_R \mbox{Tr}\left({\bf G}^2{\bf F}^{[1]2}\left(I+P_1{\bf F}^{[1]2}+P_2{\bf F}^{[2]2}+P_R{\bf G}^2\right)^{-1}\right)\right)\\
&=& \log\left(1+P_1P_R\sum_{k=1}^R\frac{f_k^{[1]2}g_k^2}{1+P_1f_k^{[1]2}+ P_2f_k^{[2]2}+P_R g_k^2}\right).
\end{eqnarray}
}
This rate is achieved with relay amplification matrix ${\bf D}(\theta=\pi/2)$,
\begin{eqnarray}
{\bf D}(\pi/2)= \gamma P_1 {\bf G}{\bf  F}^{[1]}\left(I+P_1 {\bf F}^{[1]2}+P_2 {\bf F}^{[2]2}+P_R {\bf G}^2\right)^{-1}
\end{eqnarray}
Further, if user $1$ achieves his maximum rate $C_{10}$, i.e. ${\bf D}={\bf D}(\pi/2)$, then the maximum rate that can be achieved by user $2$ is 
\begin{eqnarray}
C_2^{10} &=& \log\left(1+P_2P_R\frac{\left(\sum_{i=1}^R\frac{g_i^2f_i^{[1]}f_i^{[2]}}{1+P_1f_i^{[1]2}+P_2f_i^{[2]2}+P_Rg_i^2}\right)^2}{\sum_{i=1}^R\frac{g_i^2f_i^{[1]2}}{1+P_1f_i^{[1]2}+P_2f_i^{[2]2}+P_Rg_i^2}+P_1P_R\left(\sum_{i=1}^R\frac{g_i^2f_i^{[1]2}}{1+P_1f_i^{[1]2}+P_2f_i^{[2]2}+P_Rg_i^2}\right)^2}\right)\nonumber
\end{eqnarray}
\end{theorem}
The proof of Theorem \ref{theorem:mu1} is in Appendix \ref{app:mu1}.

By symmetry, the expression for maximum rate $C_2^{01}$ for user $2$ is obtained by switching indices $1$ and $2$ in Theorem \ref{theorem:mu1}. Note that user $2$ obtains his maximum rate when $\theta=0$, i.e., with amplification matrix ${\bf D}(0)$.

Theorem \ref{theorem:mu1} leads to an interesting observation. Notice that $C_1^{10}$, i.e.,  the maximum rate for user $1$ in  MAC$\left({\bf F}^{[1]}, P_1, {\bf F}^{[2]}, P_2, {\bf G}, \{{\bf D}\}, P_R\right)$ is not the same as the single user capacity of user $1$ if user $2$ was \emph{not} transmitting. In fact, for $P_1, P_2 >0$,
\begin{eqnarray}
C_1^{10} < C^{PTP}({\bf F}^{[1]},P_1,{\bf G},\{{\bf D}\},P_R).
\end{eqnarray}
Recall that in the conventional Gaussian MAC (i.e. without AF relays) the maximum rate that user 1 can achieve is the same as his channel capacity \emph{as if} user 2 is not transmitting. In the conventional Gaussian MAC, even though user $2$ transmits power $P_2$, user $1$ can achieve his single user capacity with power $P_1$ if the rate $R_2$ allows user $2$ to be decoded and subtracted out before user $1$ is decoded. The reason this is not true in the relay MAC is because of the power constraint at the relays. Even though user $2$ can be decoded and his signal subtracted out at the destination, it does not make the system equivalent to one where user $2$ was not transmitting at all. Transmission by user $2$ affects the power constraint at the AF relays, thereby strictly reducing user $1$'s channel capacity as compared to the case when user $2$ is silent.

Next we consider the sum rate capacity, i.e. $\mu_1=\mu_2=1$.
\subsection{Sum Rate Capacity}
\begin{theorem}\label{theorem:sumrate}
The sum rate capacity $C^{11}$ of MAC$\left({\bf F}^{[1]}, P_1, {\bf F}^{[2]}, P_2, {\bf G}, \{{\bf D}\}, P_R\right)$, i.e., a $2$ user MAC with $R$ parallel AF relays, when the user's transmit powers are $P_1$ and $P_2$, the total transmit power of all relays is $P_R$ and the relay amplification factor is optimized over all feasible ${\bf D}$:
\begin{eqnarray*}
C^{11} &=& \log\left(1+\mbox{SNR}^\star\right)\\
\mbox{SNR}^\star &=&\frac{P_R}{2}\left(P_1A_{11}+P_2A_{22}+\sqrt{\left(P_1A_{11}+P_2A_{22}\right)^2-4P_1P_2\left(A_{11}A_{22}-A_{12}^2\right)}\right)
\end{eqnarray*}
where
\begin{eqnarray*}
A_{11} &= &\sum_{k=1}^R\frac{g_k^2f_k^{[1]2}}{1+P_1f_k^{[1]2}+P_2f_k^{[2]2}+P_Rg_k^2}=\mbox{Tr}\left({\bf G}^2{\bf F}^{[1]2}\left(I+P_1{\bf F}^{[1]2}+P_2{\bf F}^{[2]2}+P_R{\bf G}^2\right)^{-1}\right)\\
A_{22} &= &\sum_{k=1}^R\frac{g_k^2f_k^{[2]2}}{1+P_1f_k^{[1]2}+P_2f_k^{[2]2}+P_Rg_k^2}=\mbox{Tr}\left({\bf G}^2{\bf F}^{[2]2}\left(I+P_1{\bf F}^{[1]2}+P_2{\bf F}^{[2]2}+P_R{\bf G}^2\right)^{-1}\right)\\
A_{12} &= &\sum_{k=1}^R\frac{g_k^2f_k^{[1]}f_k^{[2]}}{1+P_1f_k^{[1]2}+P_2f_k^{[2]2}+P_Rg_k^2}=\mbox{Tr}\left({\bf G}^2{\bf F}^{[1]}{\bf F}^{[2]}\left(I+P_1{\bf F}^{[1]2}+P_2{\bf F}^{[2]2}+P_R{\bf G}^2\right)^{-1}\right)\\
\end{eqnarray*}
The sum rate capacity is achieved with relay amplification matrix ${\bf D}(\theta=\theta^{11})$, where
\begin{eqnarray}
\theta^{11} = \tan^{-1}\left({\frac{P_RP_2A_{12}}{\mbox{SNR}^\star-P_RP_1A_{11}}}\right)
\end{eqnarray}
Further, the following rates pairs $(R_1,R_2)$ are sum rate optimal:
\begin{eqnarray*}
(R_1^{11}(2\rightarrow 1),R_2^{11}(2\rightarrow 1)) &=& \left(\log(1+\beta \mbox{SNR}^\star),\log\left(1+\frac{(1-\beta)\mbox{SNR}^\star}{1+\beta\mbox{SNR}^\star}\right)\right)\\
(R_1^{11}(1\rightarrow 2),R_2^{11}(1\rightarrow 2)) &=& \left(\log\left(1+\frac{\beta \mbox{SNR}^\star}{1+(1-\beta)\mbox{SNR}^\star}\right),\log(1+(1-\beta) \mbox{SNR}^\star)\right)\\
\mbox{where}~~~~~~ \beta &=& \frac{\mbox{SNR}^\star-P_2P_RA_{22}}{2\mbox{SNR}^\star-P_1P_RA_{11}-P_2P_RA_{22}},
\end{eqnarray*}
and the rate pairs are achieved by successive decoding in the order $2$ followed by $1$, and $1$ followed by $2$, respectively.
\end{theorem}
Proof of Theorem \ref{theorem:sumrate} is presented in Appendix \ref{app:sumrate}. Notice from Theorem \ref{theorem:sumrate} that the sum rate maximizing point is not unique. In other words the capacity region of the parallel AF relay MAC$\left({\bf F}^{[1]}, P_1, {\bf F}^{[2]}, P_2, {\bf G}, \{{\bf D}\}, P_R\right)$ is not strictly convex in the region of sum rate optimality, at least for fixed $P_1,P_2$. While the capacity region is not exactly a pentagon, it is similar to a pentagon with two smooth corners. The smooth corners arise from the different optimal ${\bf D}$ that maximize $\mu_1R_1+\mu_2R_2$ for different $\mu_1,\mu_2$. This is similar in shape to the capacity region of a MIMO Gaussian MAC where different input covariance matrices are optimal for different $\mu_1,\mu_2$ pairs. Similar to the MIMO MAC, the capacity region is a straight line around the sum rate optimal region, connecting the two different rate pairs that both achieve the sum rate capacity. The intermediate points on the line correspond to rate pairs that are achieved with time sharing of the extreme points described in Theorem \ref{theorem:sumrate}.

\subsection{Optimal Rate Pair to maximize $\mu_1R_1+\mu_2R_2$}
\begin{theorem}\label{theorem:mu1mu2}
For any $\mu_1,\mu_2$ with $\mu_1\geq\mu_2$, the weighted sum rate $\mu_1R_1+\mu_2R_2$ for the parallel AF relay MAC$\left({\bf F}^{[1]}, P_1, {\bf F}^{[2]}, P_2, {\bf G}, \{{\bf D}\}, P_R\right)$ is maximized  by the rate pair $(R_1^{\mu_1,\mu_2},R_2^{\mu_1,\mu_2})$ given by 
\begin{eqnarray}
R_1^{\mu_1,\mu_2}&=&\log(1+\mbox{SNR}_1^\star), ~~~~~~~~R_2^{\mu_1,\mu_2}~=~\log\left(1+\frac{\mbox{SNR}_2^\star}{1+\mbox{SNR}_1^\star}\right)
\end{eqnarray}
where $\mbox{SNR}_1^\star$ and $\mbox{SNR}_2^\star$ are the solutions to the following simultaneous equations:
\begin{eqnarray}
\mu_1&=&\frac{\mu_2}{\mbox{SNR}_1^\star+\mbox{SNR}_2^\star}(1+P_RP_1A_{11}+P_RP_2A_{12}/\alpha)+\frac{\mu_1-\mu_2}{1+\mbox{SNR}_1^\star}(1+P_RP_1A_{11})\\
\mu_1&=&\frac{\mu_2}{\mbox{SNR}_1^\star+\mbox{SNR}_2^\star}(1+P_RP_1A_{12}\alpha+P_RP_2A_{22})+\frac{\mu_1-\mu_2}{1+\mbox{SNR}_1^\star}(1+P_RP_1A_{12}\alpha)\\
\alpha &=& \sqrt{\frac{P_1\mbox{SNR}_1^\star}{P_2\mbox{SNR}_2^\star}}
\end{eqnarray}
\end{theorem}
Proof of Theorem \ref{theorem:mu1mu2} is provided in Appendix \ref{app:mu1mu2}. Note that for $\mu_2\geq\mu_1$ the corresponding rate pairs can be obtained by switching indices $1,2$ in Theorem \ref{theorem:mu1mu2}. While the simultaneous equations do not appear to allow a closed form solution for general $\mu_1,\mu_2$,  closed form solutions can be obtained for the special cases of $\mu_1=1,\mu_2=0$ and $\mu_1=\mu_2=1$ as discussed above, . In general, Theorem \ref{theorem:mu1mu2} allows a relatively straightforward numerical solution that is much easier than a brute force optimization over all feasible ${\bf D}$. Solving the weighted sum of rate pairs optimization problem is especially useful for rate allocation problems as well as to plot the entire capacity region of the parallel AF relay MAC.
\subsection{The Capacity Region}
\begin{figure}[h]
\centerline{\input{maccap.pstex_t}}
\caption{Capacity region of two user AF relay MAC$\left({\bf F}^{[1]}, P_1, {\bf F}^{[2]}, P_2, {\bf G}, \{{\bf D}\}, P_R\right)$}\label{fig:theta}
\end{figure}
Combining the results of Theorems \ref{theorem:maccap},\ref{theorem:mu1},\ref{theorem:sumrate} and \ref{theorem:mu1mu2} the boundary of the capacity region for the parallel AF relay MAC$\left({\bf F}^{[1]}, P_1, {\bf F}^{[2]}, P_2, {\bf G}, \{{\bf D}\}, P_R\right)$ is graphically plotted by the following algorithm. The steps are labeled with reference to Fig. \ref{fig:theta}.
\begin{itemize}
\item {A$\rightarrow$ B}: Draw a straight horizontal line from $(R_1,R_2)=(0,C_2^{01})$ to $(R_1,R_2)=(C_1^{01},C_2^{01})$.
\item {B$\rightarrow$ C}: For $\theta: 0\rightarrow\theta^{11}$
\begin{eqnarray}
{\bf D}(\theta) &=& {\bf G}\left(P_1{\bf  F}^{[1]}\sin\theta+P_2 {\bf F}^{[2]}\cos\theta\right)\left(I+P_1 {\bf F}^{[1]2}+P_2 {\bf F}^{[2]2}+P_R {\bf G}^2\right)^{-1}\\
R_1&=& \log\left(1+P_1P_R\frac{[\mbox{Tr}\left({\bf G}{\bf D}{\bf F}^{[1]}\right)]^2}{{\mbox{Tr}\left[{\bf D}^2\left(I+P_1{\bf F}^{[1]2}+P_2{\bf F}^{[2]2}+P_R{\bf G}^2\right)\right]}}+P_2P_R[\mbox{Tr}\left({\bf G}{\bf D}{\bf F}^{[2]}\right)]^2\right)\nonumber\\
R_2&=&\log\left(1+P_2P_R\frac{[\mbox{Tr}\left({\bf G}{\bf D}{\bf F}^{[2]}\right)]^2}{{\mbox{Tr}\left[{\bf D}^2\left(I+P_1{\bf F}^{[1]2}+P_2{\bf F}^{[2]2}+P_R{\bf G}^2\right)\right]}}\right)\nonumber
\end{eqnarray}
\item {C$\rightarrow$ D}: Draw a straight line from $(R_1^{11}(1\rightarrow 2),R_2^{11}(1\rightarrow 2)$ to $(R_1^{11}(2\rightarrow 1),R_2^{11}(2\rightarrow 1)$.
\item {D$\rightarrow$ E}: For $\theta: \theta^{11}\rightarrow\mbox{sgn}(\theta^{11})\frac{\pi}{2}$
\begin{eqnarray}
{\bf D}(\theta) &=& {\bf G}\left(P_1{\bf  F}^{[1]}\sin\theta+P_2 {\bf F}^{[2]}\cos\theta\right)\left(I+P_1 {\bf F}^{[1]2}+P_2 {\bf F}^{[2]2}+P_R {\bf G}^2\right)^{-1}\\
R_1&=&\log\left(1+P_1P_R\frac{[\mbox{Tr}\left({\bf G}{\bf D}{\bf F}^{[1]}\right)]^2}{{\mbox{Tr}\left[{\bf D}^2\left(I+P_1{\bf F}^{[1]2}+P_2{\bf F}^{[2]2}+P_R{\bf G}^2\right)\right]}}\right)\nonumber\\
R_2&=& \log\left(1+P_2P_R\frac{[\mbox{Tr}\left({\bf G}{\bf D}{\bf F}^{[2]}\right)]^2}{{\mbox{Tr}\left[{\bf D}^2\left(I+P_1{\bf F}^{[1]2}+P_2{\bf F}^{[2]2}+P_R{\bf G}^2\right)\right]}}+P_1P_R[\mbox{Tr}\left({\bf G}{\bf D}{\bf F}^{[1]}\right)]^2\right)\nonumber
\end{eqnarray}
\item {E$\rightarrow$ F}: Draw a straight vertical line from $(R_1,R_2)=(C_1^{10},C_2^{10})$ to $(R_1,R_2)=(C_1^{10},0)$.
\end{itemize}
Fig. \ref{fig:theta} shows the typical shape of the capacity region. Optimal relay amplification factor is indicated on the figure in terms of the parameter $|\theta|$.  $\theta$ is equal to zero between points $A$ and $B$, it changes from $0$ to $\theta^{11}$ (as defined in Theorem \ref{theorem:sumrate}) in the curved portion from points $B$ to $C$. $\theta$ is constant at $\theta^{11}$ between points $C$ to $D$. $\theta$ changes from $\theta^{11}$ to $\mbox{sgn}(\theta^{11})\pi/2$ as we traverse the boundary from $E$, and it is again constant at $\theta=\pi/2$ from the point $E$ to point $F$. The coordinates of the points $A,B,C,D,E,F$ are all known in closed form as given by the preceding results in this section. Point $A'$ outside the capacity region indicates the maximum rate of user $2$ with power $P_2$ if user $1$ is not transmitting, i.e. $P_1=0$. This is strictly higher than point $A$ which corresponds to the maximum rate of user $2$ with power $P_2$ when user $1$ is sending constant symbols ($R_1=0$) that can be subtracted out at the receiver. As explained earlier, unlike the conventional Gaussian MAC, user $1$ sending constant symbols with power $P_1$ is not identical to user $1$ silent, even though the constant symbols can be subtracted by the destination. This is because transmission by user $1$ affects the relay power constraint. Notice that AF relays are only allowed to scale the input, i.e. they cannot subtract the constant signal from user $1$. Similarly, point $F'$ indicates the maximum rate of user $1$ if user $2$ is not transmitting, which is  higher than the maximum rate of user $1$ when user $2$ is transmitting (point $F$).

\section{Duality Relationships in Parallel AF Relay Networks}
In this section we examine the duality of the parallel AF relay MAC and BC. In the previous section we obtained the capacity region of the parallel AF MAC. Obtaining an equally explicit characterization for the parallel AF BC is much harder because, as we will see, without time-sharing among different ${\bf D}$ the BC capacity region is often non-convex. The notion of duality is particularly useful in such a scenario. Knowing the capacity region of the AF relay MAC would enable us to find the capacity region of the AF relay BC if we could obtain a duality relationship between the two. Establishing such a duality is the goal of this section.
\subsection{Duality of Parallel AF Relay MAC and BC}\label{sec:macbcdual}
The following two theorems establish the duality relationship between the parallel AF relay MAC and BC.
\begin{theorem}\label{theorem:macbc}
\begin{eqnarray}
\mathcal{C}^{MAC}({\bf F}^{[1]}, P_1, {\bf F}^{[2]}, P_2, {\bf G}, {\bf D}, P_R)&\subset&\mathcal{C}^{BC}\left({\bf G}, P_R, {\bf F}^{[1]}, {\bf F}^{[2]}, \kappa{\bf D}, P_1+P_2\right)
\end{eqnarray}
Given any relay amplification matrix ${\bf D}$ that satisfies the power constraint on the parallel AF relay multiple access channel MAC$({\bf F}^{[1]}, P_1, {\bf F}^{[2]}, P_2, {\bf G}, {\bf D}, P_R)$, there exists a dual parallel AF relay broadcast channel BC$\left({\bf G}, P_R, {\bf F}^{[1]}, {\bf F}^{[2]}, \kappa{\bf D}, P_1+P_2\right)$ such that any rate pair $(R_1, R_2)$ that can be achieved on MAC$({\bf F}^{[1]}, P_1, {\bf F}^{[2]}, P_2, {\bf G}, {\bf D}, P_R)$ can also be achieved on BC$\left({\bf G}, P_R, {\bf F}^{[1]}, {\bf F}^{[2]}, \kappa\{{\bf D}\}, P\right)$. $\kappa$ is chosen to satisfy the relay sum power constraint on BC$\left({\bf G}, P_R, {\bf F}^{[1]}, {\bf F}^{[2]}, \kappa{\bf D}, P_1+P_2\right)$.
\end{theorem}
\begin{theorem}\label{theorem:macbc2}
\begin{eqnarray}
\mathcal{C}^{BC}\left({\bf G}, P_R, {\bf F}^{[1]}, {\bf F}^{[2]},{\bf D}, P\right) = \cup_{P_1,P_2\geq 0, P_1+P_2=P}~~~\mathcal{C}^{MAC}({\bf F}^{[1]}, P_1, {\bf F}^{[2]}, P_2, {\bf G},  \kappa{\bf D}, P_R)
\end{eqnarray}
Given any relay amplification matrix ${\bf D}$ that satisfies the power constraint on the parallel AF relay broadcast channel BC$\left({\bf G}, P_R, {\bf F}^{[1]}, {\bf F}^{[2]}, {\bf D}, P\right)$ and given a rate pair $(R_1,R_2)$ that is achievable on this parallel AF relay broadcast channel, there exist $P_1, P_2\geq 0$ such that $P_1+P_2=P$ and a dual multiple access channel MAC$({\bf F}^{[1]}, P_1, {\bf F}^{[2]}, P_2, {\bf G}, \kappa{\bf D}, P_R)$ such that the rate pair $R_1, R_2$ is achievable on MAC$({\bf F}^{[1]}, P_1, {\bf F}^{[2]}, P_2, {\bf G}, \kappa{\bf D}, P_R)$. $\kappa$ is chosen to satisfy the relay sum power constraint on MAC$({\bf F}^{[1]}, P_1, {\bf F}^{[2]}, P_2, {\bf G}, {\bf D}, P_R)$.
\end{theorem}
The proof of Theorems \ref{theorem:macbc} and \ref{theorem:macbc2} is presented in Appendix \ref{app:macbc}. 



Note that the duality relationship is a strong duality in the sense that the parallel AF relay MAC and BC are duals not only for the optimal relay amplification matrix ${\bf D}$ but also for \emph{any} feasible ${\bf D}$. This is an equally strong result as for the point to point case. Moreover, as in the point to point case, note that the powers of the relays and the transmitter are switched in the dual channel
\subsection{Capacity Region of the Parallel AF Relay BC}
\begin{figure}[h]
\epsfig{figure=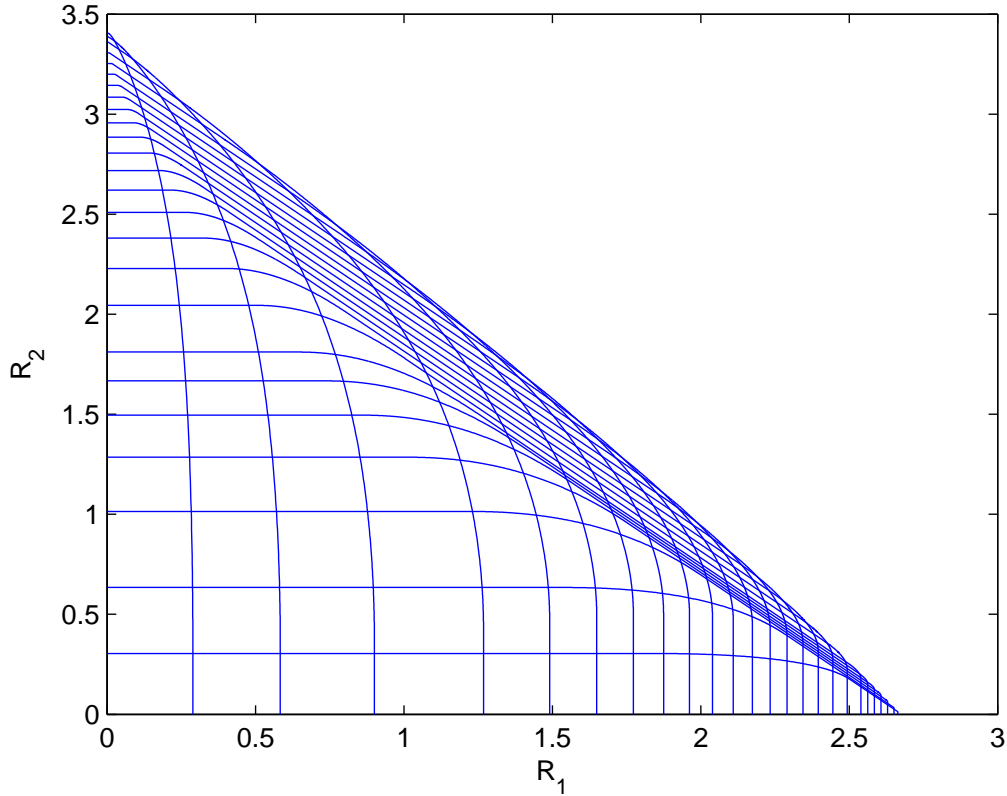, width = 6in, height = 4.5in}
\caption{BC capacity region as the union of MAC capacity regions}\label{fig:bcmac}
\end{figure}
The duality relationship described in the previous section provides a method to compute the capacity region of the parallel AF broadcast channel. As shown in Fig. \ref{fig:bcmac} the BC capacity region may be found simply as the union of the MAC capacity regions. Interestingly, while the MAC capacity region for fixed $P_1,P_2$ was found to be convex in all numerical results, the BC capacity region is often a non-convex region. Note that the non-convex nature of the capacity region is because we have not considered time sharing between relay amplification factors. Indeed with time sharing between different relay factors, the capacity region can always be made convex. However, if time sharing between relay amplification factors is not allowed, the union of the MAC capacity regions for all different combinations of $P_1+P_2=P$, and hence the BC capacity region is often non-convex. While we have not shown the convexity of the MAC capacity region for any fixed $P_1,P_2$, we expect it to be always convex. This hypothesis is supported by the fact that for fixed $P_1, P_2$ the MAC sum capacity is maximized by a unique relay amplification ${\bf D}$ matrix, which indicates time sharing between different ${\bf D}$ may not be necessary to achieve any given point on the MAC capacity region.
\subsection{Generalized AF MAC-BC Duality with Multiple Antenna Relays and Multiple Hops}
\begin{figure}[h]
\centerline{\input{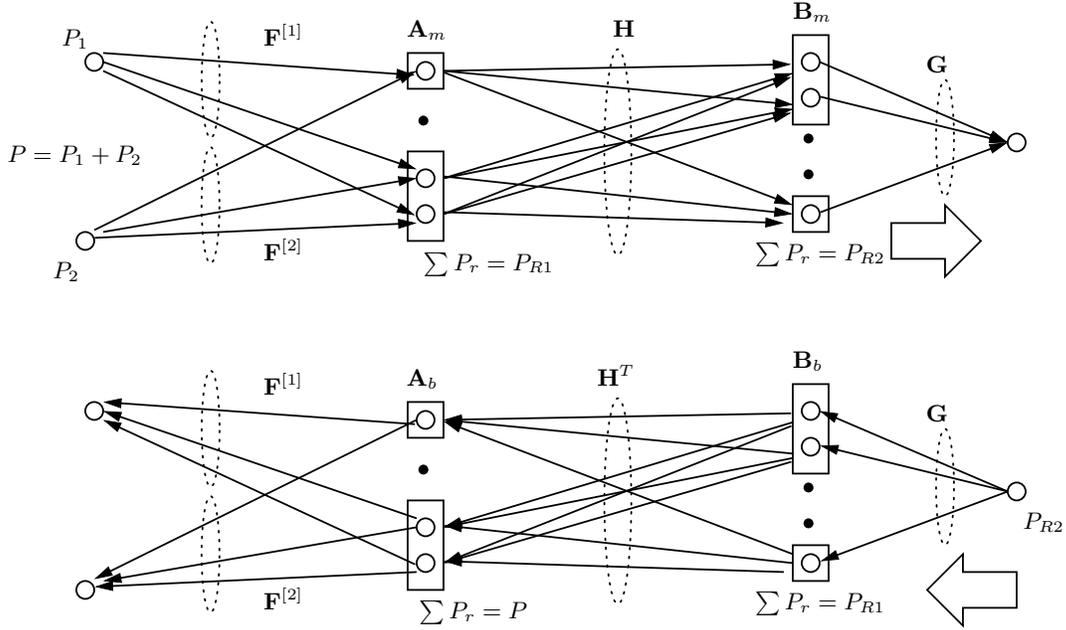}}
\caption{3-hop MAC-BC duality with distributed multiple antenna AF relays}\label{fig:multaf}
\end{figure}

The AF relay MAC-BC duality shown in section \ref{sec:macbcdual} is not limited to single antenna relays or two hop networks with parallel relays. The same duality relationship holds when some of the relays have multiple antennas. The duality also holds when more than $2$ hops are considered. Thus, the MAC-BC duality holds for AF relay networks whether they are purely parallel (two hop), purely serial (multiple hops with a single relay at each hop), or several parallel AF relay clusters connected in series to form a multihop relay network. It holds whether the relays are distributed with a single antenna at each relay or they are able to cooperate as as a multiple antenna node. The dual network in general is defined as the network where the MAC source nodes become the BC destination nodes and the MAC destination node becomes the BC source node. The power constraints are changed so that the link on each hop is subject to the same transmit power as in the dual network. Consider the two hop MAC. The source to relay channel (let us call it link $1$),  is subject to total transmit power $P_1+P_2 = P$ while the relay destination channel (link $2$) is subject to total transmit power $P_R$. In the dual BC, the relay power becomes $P$ so that the relay to destination channel (link $1$) is still associated with power $P$. The source power on the dual BC is $P_R$ so that link $2$ is still associated with power $P_R$. In the general case, consider an $m$ hop MAC where successive hops (starting with the source) are associated with sum power constraints $P_1, P_2, \cdots, P_m$ respectively. Then in the dual BC, starting with the source, the power constraints will be $P_m, P_{m-1}, \cdots, P_1$ so that $P_1$ is the power of the last set of parallel relays.

Since the proof technique is the same in all cases, we illustrate the proof with an example of a three hop relay network when multiple antennas may be present at some or all of the relay nodes.  Multiple antenna AF relays can forward any linear transformation of their inputs. In other words, the amplification factor for a multiple antenna AF relay node can be any matrix with the same dimensions as the number of antennas at the relay. The overall amplification factor of each group of relays corresponding to a hop of the relay network can be represented then as block diagonal matrix. For the three hop example in this section, we use symbols ${\bf A}, {\bf B}$ to indicate the relay amplification block matrices for the two relay hops, with the blocks representing the scaling matrices of multiple antenna relays. As a matter of notation, we denote the scaling matrices on the MAC as ${\bf A}_m, {\bf B}_m$ while we denote the scaling matrix on the BC as ${\bf A}_b, {\bf B}_b$. Let ${\bf 1}_{R\times 1}$ be a column vector containing all ones. Note that while we assume $R$ is the number of relays at each hop for simplicity, the same proof applies even if the number of relays at each hop is different. We will use the following definitions to simplify the notation for this section:
{\allowdisplaybreaks
\begin{eqnarray}
\overline{\bf G} &=& {\bf G}{\bf 1}_{R\times 1}\\
\overline{\bf F}^{[1]} &=& {\bf F}^{[1]}{\bf 1}_{R\times 1}\\
\overline{\bf F}^{[2]} &=& {\bf F}^{[2]}{\bf 1}_{R\times 1}\\
\overline{\bf N}_{R,1} &=& {\bf N}_{R,1}{\bf 1}_{R\times 1}
\end{eqnarray}
}
Thus, $\overline{\bf G},\overline{\bf F}^{[1]},\overline{\bf F}^{[2]},\overline{\bf N}_{R,1}$ are column vectors. Symbol ${\bf H}$ indicates the MIMO channel between the two relay stages. $P_{R1}$ and $P_{R2}$ are the transmit powers of the two relay stages. The precise definitions of these symbols will be clear from Fig. \ref{fig:multaf} and the following analysis.

The output signals for the MAC and BC are:
{\allowdisplaybreaks
\begin{eqnarray}
\mbox{MAC}&& y= \overline{\bf G}^T{\bf B}_m\left({\bf H}{\bf A}_m\left( \overline{\bf F}^{[1]}x_1+ \overline{\bf F}^{[2]}x_2 +  \overline{\bf N}_{R,1}\right) + {\bf N}_{R,2}\right) +n\\
\mbox{BC}&&\left\{\begin{array}{ccc} y_1 &=&\overline{\bf F}^{[1]T}{\bf A}_b\left({\bf H}^T{\bf B}_b\left(\overline{\bf G}~ x+\overline{\bf N}_{R,1}\right)+{\bf N}_{R,2}\right)+n_1,\\ y_2 &=&\overline{\bf F}^{[2]T}{\bf A}_b\left({\bf H}^T{\bf B}_b\left(\overline{\bf G}~ x+\overline{\bf N}_{R,1}\right)+{\bf N}_{R,2}\right)+n_2. \end{array}\right.
\end{eqnarray}
}
 With noise power normalized to unity the MAC-BC are characterized as:
{\allowdisplaybreaks
\begin{eqnarray}
\mbox{MAC}&& y= \frac{\overline{\bf G}^T{\bf B}_m{\bf H}{\bf A}_m\overline{\bf F}^{[1]}x_1+\overline{\bf G}^T{\bf B}_m{\bf H}{\bf A}_m\overline{\bf F}^{[2]}x_2}{\sqrt{1+\overline{\bf G}^T{\bf B}_m{\bf B}_m^T\overline{\bf G}+\overline{\bf G}^T{\bf B}_m{\bf H}{\bf A}_m{\bf A}_m^T{\bf H}^T{\bf B}_m^T\overline{\bf G}}}~~+n, ~~~~~~~\mbox{E}[x_1^2]=P_1, \mbox{E}[x_2^2]=P_2 \nonumber\\
&&\mbox{Relay Power Constraints for MAC:} ~~~~~~~~~~~~~~~~~~~~~~~~~~~~~~~~~~~~~~~~~~~~~~~~~~~~~~~~~~~~~~~~~~~~~~~~~~~~~~~~~~~~~~~~\nonumber\\
&&P_1 \mbox{Tr}\left({\bf A}_m \overline{\bf F}^{[1]}\overline{\bf F}^{[1]T}{\bf A}^T_m\right)+P_2 \mbox{Tr}\left({\bf A}_m \overline{\bf F}^{[2]}\overline{\bf F}^{[2]T}{\bf A}^T_m\right)+\mbox{Tr}({\bf A}_m{\bf A}^T_m)=P_{R1}.\nonumber\\
&&P_1 \mbox{Tr}\left({\bf B}_m{\bf H}{\bf A}_m \overline{\bf F}^{[1]}\overline{\bf F}^{[1]T}{\bf A}^T_m{\bf H}^T{\bf B}_m^T\right)+P_2  \mbox{Tr}\left({\bf B}_m{\bf H}{\bf A}_m \overline{\bf F}^{[2]}\overline{\bf F}^{[2]T}{\bf A}^T_m{\bf H}^T{\bf B}_m^T\right)\nonumber\\
&&~~~~~~~~~~~~~~~~~~~~~~~~~~~~~~~~~~~~+\mbox{Tr}\left({\bf B}_m{\bf H}{\bf A}_m {\bf A}^T_m{\bf H}^T{\bf B}_m^T\right)+\mbox{Tr}({\bf B}_m{\bf B}^T_m)=P_{R2}.\nonumber\\
\mbox{BC}&&\left\{\begin{array}{ccc} y_1 &=&\frac{\overline{\bf F}^{[1]T}{\bf A}_b{\bf H}^T{\bf B}_b\overline{\bf G}}{\sqrt{1+\overline{\bf F}^{[1]T}{\bf A}_b{\bf A}_b^T\overline{\bf F}^{[1]}+\overline{\bf F}^{[1]T}{\bf A}_b{\bf H}^T{\bf B}_b{\bf B}_b^T{\bf H}{\bf A}_b^T\overline{\bf F}^{[1]}}}~~ x+n_1,~~\mbox{E}[x^2]=P_{R2}\\ y_2 &=& \frac{\overline{\bf F}^{[2]T}{\bf A}_b{\bf H}^T{\bf B}_b\overline{\bf G}}{\sqrt{1+\overline{\bf F}^{[2]T}{\bf A}_b{\bf A}_b^T\overline{\bf F}^{[2]}+\overline{\bf F}^{[2]T}{\bf A}_b{\bf H}^T{\bf B}_b{\bf B}_b^T{\bf H}{\bf A}_b^T\overline{\bf F}^{[2]}}}~~ x+n_2.\end{array}\right.\nonumber\\
&&\mbox{Relay Power Constraints for BC:} ~~~~~~~~~~~~~~~~~~~~~~~~~~~~~~~~~~~~~~~~~~~~~~~~~~~~~~~~~~~~~~~~~~~~~~~~~~~~~~~~~~~~~~~~\nonumber\\
&&P_{R2}\mbox{Tr}\left({\bf B}_b\overline{\bf G}\overline{\bf G}^T{\bf B}_b^T\right) +\mbox{Tr}({\bf B}_b{\bf B}_b^T)=P_{R1}\nonumber\\
&&P_{R2}\mbox{Tr}\left({\bf A}_b{\bf H}^T{\bf B}_b\overline{\bf G}\overline{\bf G}^T{\bf B}_b^T{\bf H}{\bf A}_b^T\right) +\mbox{Tr}\left({\bf A}_b{\bf H}^T{\bf B}_b{\bf B}_b^T{\bf H}{\bf A}_b^T\right)+\mbox{Tr}({\bf A}_b{\bf A}_b^T)=P=P_1+P_2\nonumber
\end{eqnarray}
}
The following theorem states the duality result in this case.
\begin{theorem}\label{theorem:macbcgeneral}
For the 3-hop distributed multiple antenna AF relay MAC and BC described above and depicted in Fig. \ref{fig:multaf},
\begin{eqnarray}
\mathcal{C}^{MAC}({\bf F}^{[1]}, P_1, {\bf F}^{[2]}, P_2, {\bf H}, {\bf A}, P_{R1},{\bf G},{\bf B},P_{R2})&\subset&\mathcal{C}^{BC}\left({\bf G}, P_{R2}, \kappa_1{\bf B}^T, P_{R1}, {\bf H}^T, {\bf F}^{[1]}, {\bf F}^{[2]}, \kappa_2{\bf A}^T, P_1+P_2\right)\nonumber
\end{eqnarray}
and
\begin{eqnarray}
&&\mathcal{C}^{BC}\left({\bf G}, P_{R2}, \kappa_1{\bf B}^T, P_{R1}, {\bf H}^T, {\bf F}^{[1]}, {\bf F}^{[2]}, \kappa_2{\bf A}^T, P_1+P_2\right)\nonumber\\
&&~~~~~~~= \cup_{P_1,P_2\geq 0, P_1+P_2=P}~~~\mathcal{C}^{MAC}({\bf F}^{[1]}, P_1, {\bf F}^{[2]}, P_2, {\bf H}, {\bf A}, P_{R1},{\bf G},{\bf B},P_{R2})
\end{eqnarray}
\end{theorem}
Notice that duality holds when ${\bf A}_m=\kappa_1{\bf A}_b^T$ and ${\bf B}_m=\kappa_2{\bf B}_b^T$. The constants are determined by the relay power constraints.
\section{Conclusion}
We explored the capacity and duality aspects of AF relay networks. The MAC-BC duality known for conventional one hop Gaussian channel was found to be applicable to two hop communication over AF relay networks where some of the relays may have multiple antennas. A unique aspect of the AF relay MAC-BC duality is that the powers of the transmitter and the relays are switched in the dual network. With distributed single antenna relay nodes we determined the optimal relay scaling factors for the entire capacity region of the relay multiple access channel. Closed form expressions were found for the sum rate and individual maximum rates while simultaneous equations were found that can be solved to determine any rate pair on the boundary of the relay MAC. The capacity region of the relay BC was evaluated using duality as the union of the relay MAC capacity regions over different power splits between user 1 and user 2 while keeping the total power constant and equal to the total relay transmit power on the BC.

We conclude with a word about the generality of the results. While we focus on real channel, signal, noise and amplification factors, the duality results and the associated proofs extend with almost no change to complex channels, signals and amplification factors as well. While the duality can be extended to multiple users as well, we have focused on the two user case to avoid the cumbersome notational aspects of considering multiple users in addition to multiple relays multiple hops and multiple antennas. The capacity characterization of the relay MAC for the complex case will also require phase optimizations of the relay scaling factors which may be non-trivial. Finally, we conjecture the duality results will hold for AF relay MIMO MAC-BC as well. However MIMO channels may require more work to establish equivalence of capacity regions as the structure of the transmitted signal covariance matrix on the dual MAC and BC will be different.

\appendix
\section{Appendix}
\subsection{Proof of Theorem \ref{theorem:maccap}}\label{proof:maccap}
Consider the normalized MAC characterization
\begin{eqnarray}
y'&=& \frac{\sum_{k=1}^R g_k d_k f_k^{[1]}}{\sqrt{1+\sum_{k=1}^Rd_k^2g_k^2}}~~x_1 + \frac{\sum_{k=1}^R g_k d_k f_k^{[2]}}{\sqrt{1+\sum_{k=1}^Rd_k^2g_k^2}}~~x_2+n'\\
y'&=&\sqrt{\frac{\mbox{SNR}_1}{P_1}}~~{x_1} + \sqrt{\frac{\mbox{SNR}_2}{P_2}}~~\frac{x_2}{P_2}+n'
\end{eqnarray}
where $\sqrt{\mbox{SNR}_1} =  \sqrt{P_1}\frac{\sum_{k=1}^R g_k d_k f_k^{[1]}}{1+\sum_{k=1}^Rd_k^2g_k^2}$ and $\sqrt{\mbox{SNR}_2} = \sqrt{P_2}\frac{\sum_{k=1}^R g_k d_k f_k^{[2]}}{1+\sum_{k=1}^Rd_k^2g_k^2}$ have been defined as such to allow compact notation. 

Characterizing the capacity region of a two user MAC is equivalent to an optimization problem where a weighted sum of rates is maximized, i.e. $\max \mu_1 R_1 +\mu_2 R_2$. Without loss of generality let us assume $\mu_1\geq\mu_2$. For every choice of the relay amplification matrix ${\bf D}$ we obtain a scalar Gaussian MAC for which the capacity region is a pentagon and this rate pair is maximized for the corner point:
\begin{eqnarray}
R_1&=& \log(1+\mbox{SNR}_1), R_2 =\log\left(1+\frac{\mbox{SNR}_2}{1+\mbox{SNR}_1}\right)\label{eq:R1R2def}
\end{eqnarray}

Thus, we need to solve the following optimization problem:
\begin{eqnarray}
\max_{d_1,\cdots,d_R}\mu_1R_1+\mu_2 R_2 &=& \max_{\bf D}\mu_1\log(1+\mbox{SNR}_1)+\mu_2\log\left(1+\frac{\mbox{SNR}_2}{1+\mbox{SNR}_1}\right)\\
&=&\max_{\bf D}(\mu_1-\mu_2)\log(1+\mbox{SNR}_1)+\mu_2\log(1+\mbox{SNR}_1+\mbox{SNR}_2)\\
&=&\max_{\bf D}\mu_1'\log(1+\mbox{SNR}_1)+\mu_2\log(1+\mbox{SNR}_1+\mbox{SNR}_2)
\end{eqnarray}
where $\mu_2$ and  $\mu_1'=\mu_1-\mu_2$ are both positive. Thus, we need to maximize 
\begin{eqnarray}
\max_{\bf D}(1+\mbox{SNR}_1)^{\mu_1'}(1+\mbox{SNR}_1+\mbox{SNR}_2)^{\mu_2}
\end{eqnarray}
subject to the power constraint (\ref{eq:powermac}).

We start with the Lagrangian formulation
\begin{eqnarray}
L({\bf D},\lambda) &=& \left(1+\frac{P_1(\sum_{k=1}^Rd_kg_kf_k^{[1]})^2}{1+\sum_{k=1}^Rd_k^2g_k^2}\right)^{\mu_1'} \left(1+\frac{P_1(\sum_{k=1}^Rd_kg_kf_k^{[1]})^2+P_2(\sum_{k=1}^Rd_kg_kf_k^{[2]})^2}{1+\sum_{k=1}^Rd_k^2g_k^2}\right)^{\mu_2}\nonumber\\
&&~~~~~~~~~~~-\lambda\left[\sum_{k=1}^Rd_k^2\left(1+P_1f_k^{[1]2}+P_2f_k^{[2]2}\right)-P_R\right]
\end{eqnarray}
Setting the derivative $\frac{\partial L({\bf D},\lambda) }{\partial d_i}$ to zero yields the KKT characterization of optimal $d_i$:
{\allowdisplaybreaks
\begin{eqnarray}
&&(1+\mbox{SNR}_1^\star)^{\mu_1'}(1+\mbox{SNR}_1^\star+\mbox{SNR}_2^\star)^{\mu_2-1}\left[\frac{2P_1(\sum_{k=1}^Rd_kg_kf_k^{[1]})g_if_i^{[1]}+2P_2(\sum_{k=1}^Rd_kg_kf_k^{[2]})g_if_i^{[2]}}{1+\sum_{k=1}^Rd_k^2g_k^2}\right.\nonumber\\
&&-\left.\frac{P_1(\sum_{k=1}^Rd_kg_kf_k^{[1]})^2+P_2(\sum_{k=1}^Rd_kg_kf_k^{[2]})^2}{(1+\sum_{k=1}^Rd_k^2g_k^2)^2}2g_i^2d_i\right]+(1+\mbox{SNR}_1^\star)^{\mu_1'-1}(1+\mbox{SNR}_1^\star+\mbox{SNR}_2^\star)^{\mu_2}\nonumber\\
&&\left[\frac{2P_1(\sum_{k=1}^Rd_kg_kf_k^{[1]})g_if_i^{[1]}}{1+\sum_{k=1}^Rd_k^2g_k^2}-\frac{P_1(\sum_{k=1}^Rd_kg_kf_k^{[1]})^2}{(1+\sum_{k=1}^Rd_k^2g_k^2)^2}2g_i^2d_i\right]=\lambda 2d_i\left(1+P_1f_i^{[1]2}+P_2f_i^{[2]2}\right)
\end{eqnarray}
}
Absorbing the constants into $\lambda$ we have
{\allowdisplaybreaks
\begin{eqnarray}
&&\mu_2(1+\mbox{SNR}_1^\star)\left[P_1(\sum_{k=1}^Rd_kg_kf_k^{[1]})g_if_i^{[1]}+P_2(\sum_{k=1}^Rd_kg_kf_k^{[2]})g_if_i^{[2]}-(\mbox{SNR}_1^\star+\mbox{SNR}_2^\star)g_i^2d_i\right]\nonumber\\
&&+\mu_1'(1+\mbox{SNR}_1^\star+\mbox{SNR}_2^\star)\left[P_1(\sum_{k=1}^Rd_kg_kf_k^{[1]})g_if_i^{[1]}-\mbox{SNR}_1^\star g_i^2d_i\right]=\lambda d_k\left(1+P_1f_i^{[1]2}+P_2f_i^{[2]2}\right)\nonumber\\
&\Rightarrow& \mu_2(1+\mbox{SNR}_1^\star)\left[P_1(\sum_{k=1}^Rd_kg_kf_k^{[1]})^2+P_2(\sum_{k=1}^Rd_kg_kf_k^{[2]})^2-(\mbox{SNR}_1^\star+\mbox{SNR}_2^\star)\sum_{k=1}^R g_k^2d_k^2\right]\nonumber\\
&&~~~~~~~~~~+\mu_1'(1+\mbox{SNR}_1^\star+\mbox{SNR}_2^\star)\left[P_1(\sum_{k=1}^Rd_kg_kf_k^{[1]})^2-\mbox{SNR}_1^\star \sum_{k=1}^R g_k^2d_k^2\right]=\lambda P_R\\
&\Rightarrow&\mu_2(1+\mbox{SNR}_1^\star)(\mbox{SNR}_1^\star+\mbox{SNR}_2^\star)+\mu_1'(1+\mbox{SNR}_1^\star+\mbox{SNR}_2^\star)\mbox{SNR}_1^\star=\lambda P_R\\
&\Rightarrow&d_i\left[\frac{\mu_2(1+\mbox{SNR}_1^\star)(\mbox{SNR}_1^\star+\mbox{SNR}_2^\star)+\mu_1'(1+\mbox{SNR}_1^\star+\mbox{SNR}_2^\star)\mbox{SNR}_1^\star}{P_R}\left(1+P_1f_i^{[1]2}+P_2f_i^{[2]2}\right)\right.\nonumber\\
&& \left.+\mu_2(1+\mbox{SNR}_1^\star)(\mbox{SNR}_1^\star+\mbox{SNR}_2^\star)g_i^2\right]=\mu_2(1+\mbox{SNR}_1^\star)\left[P_1(\sum_{k=1}^Rd_kg_kf_k^{[1]})g_if_i^{[1]}+P_2(\sum_{k=1}^Rd_kg_kf_k^{[2]})g_if_i^{[2]}\right]\nonumber\\
&&~~~~~~~~~~~~~~~~~~~~~~~~~~~~+\mu_1'(1+\mbox{SNR}_1^\star+\mbox{SNR}_2^\star)P_1(\sum_{k=1}^Rd_kg_kf_k^{[1]})g_if_i^{[1]}\\
&\Rightarrow& d_i\left[\frac{\mu_2(1+\mbox{SNR}_1^\star)(\mbox{SNR}_1^\star+\mbox{SNR}_2^\star)+\mu_1'(1+\mbox{SNR}_1^\star+\mbox{SNR}_2^\star)(\mbox{SNR}_1^\star)}{P_R}(1+P_1f_i^{[1]2}+P_2f_i^{[2]2}+P_Rg_i^2)\right]\nonumber\\
&&=\mu_2(1+\mbox{SNR}_1^\star)\left[P_1(\sum_{k=1}^Rd_kg_kf_k^{[1]})g_if_i^{[1]}+P_2(\sum_{k=1}^Rd_kg_kf_k^{[2]})g_if_i^{[2]}\right]+\nonumber\\
&&~~~~~~~~~~~~~~~~~~~~~~~~~~~~~\mu_1'(1+\mbox{SNR}_1^\star+\mbox{SNR}_2^\star)P_1(\sum_{k=1}^Rd_kg_kf_k^{[1]})g_if_i^{[1]}\\
&\Rightarrow& d_i \frac{\mu_2(1+\mbox{SNR}_1^\star)(\mbox{SNR}_1^\star+\mbox{SNR}_2^\star)+\mu_1'(1+\mbox{SNR}_1^\star+\mbox{SNR}_2^\star)(\mbox{SNR}_1^\star)}{P_R}=\nonumber\\
&&[\mu_1'(1+\mbox{SNR}_1^\star+\mbox{SNR}_2^\star)+\mu_2(1+\mbox{SNR}_1^\star)]P_1(\sum_{k=1}^Rd_kg_kf_k^{[1]})\frac{g_if_i^{[1]}}{1+P_1f_i^{[1]2}+P_2f_i^{[2]2}+P_Rg_i^2}\nonumber\\
&&+\mu_2(1+\mbox{SNR}_1^\star)P_2(\sum_{k=1}^Rd_kg_kf_k^{[2]})\frac{g_if_i^{[2]}}{1+P_1f_i^{[1]2}+P_2f_i^{[2]2}+P_Rg_i^2} \label{eq:key}\\
&\Rightarrow& {\bf D}={\bf G}\left(c_1 P_1{\bf  F}^{[1]}+c_2 P_2 {\bf F}^{[2]}\right)\left(I+P_1 {\bf F}^{[1]2}+P_2 {\bf F}^{[2]2}+P_R {\bf G}^2\right)^{-1} ~~~~\mbox{where}~~c_1,c_2\geq 0,\\
&\Rightarrow& {\bf D}=\sqrt{c_1^2+c_2^2}{\bf G}\left(\frac{c_1}{\sqrt{c_1^2+c_2^2}} P_1{\bf  F}^{[1]}+\frac{c_2}{\sqrt{c_1^2+c_2^2}} P_2 {\bf F}^{[2]}\right)\left(I+P_1 {\bf F}^{[1]2}+P_2 {\bf F}^{[2]2}+P_R {\bf G}^2\right)^{-1} 
\end{eqnarray}
}
Defining $\gamma=\sqrt{c_1^2+c_2^2}$ and $\tan\theta = c_1/c_2$, the optimal ${\bf D}$ for maximizing $\mu_1 R_1+\mu_2 R_2$ can be expressed in the form:
\begin{eqnarray}
{\bf D}= \gamma {\bf G}\left(P_1{\bf  F}^{[1]}\sin\theta+P_2 {\bf F}^{[2]}\cos\theta\right)\left(I+P_1 {\bf F}^{[1]2}+P_2 {\bf F}^{[2]2}+P_R {\bf G}^2\right)^{-1}.
\end{eqnarray}
Note that the constant $\gamma$ can be evaluated from the power constraint as in Theorem \ref{theorem:maccap}.\hfill\QED

\subsection{Proof of Theorem \ref{theorem:mu1}}\label{app:mu1}
We are interested in the case $\mu_1=1,\mu_2=0$. Substituting these values into the optimal ${\bf D}$ characterization of (\ref{eq:key}) we have
\begin{eqnarray}
&& d_i \frac{\mbox{SNR}_1^\star}{P_R}=P_1(\sum_{k=1}^Rd_kg_kf_k^{[1]})\frac{g_if_i^{[1]}}{1+P_1f_i^{[1]2}+P_2f_i^{[2]2}+P_Rg_i^2}\label{eq:cor2}
\end{eqnarray}
which corresponds to $\theta=\pi/2$ and leads to
\begin{eqnarray}
\mbox{SNR}_1^\star = P_1P_R\sum_{i=1}^R\frac{g_i^2f_i^{[1]2}}{1+P_1f_i^{[1]2}+P_2f_i^{[2]2}+P_Rg_i^2}
\end{eqnarray}
which gives $C^{10}_1=\log(1+\mbox{SNR}_1^\star)$.
In order to find $C^{10}_2$, we need $\mbox{SNR}^\star_2$. From the definition,
\begin{eqnarray}
\frac{\mbox{SNR}_2^\star}{\mbox{SNR}_1^\star}=\frac{P_2(\sum_{k=1}^Rd_kg_kf_k^{[2]})^2}{P_1(\sum_{k=1}^Rd_kg_kf_k^{[1]})^2}
\end{eqnarray}
From (\ref{eq:cor2}) we have
\begin{eqnarray}
&&\frac{(\sum_{k=1}^Rd_kg_kf_k^{[2]})}{(\sum_{k=1}^Rd_kg_kf_k^{[1]})} {\mbox{SNR}_1^\star}{}=P_1P_R\sum_{i=1}^R\frac{g_i^2f_i^{[1]}f_i^{[2]}}{1+P_1f_i^{[1]2}+P_2f_i^{[2]2}+P_Rg_i^2}\\
&\Rightarrow&\mbox{SNR}_2^\star=\frac{(\sum_{k=1}^Rd_kg_kf_k^{[2]})^2}{(\sum_{k=1}^Rd_kg_kf_k^{[1]})^2} {\mbox{SNR}_1^\star}\frac{P_2}{P_1}\\
&\Rightarrow&\mbox{SNR}_2^\star=P_2P_R\frac{\left(\sum_{i=1}^R\frac{g_i^2f_i^{[1]}f_i^{[2]}}{1+P_1f_i^{[1]2}+P_2f_i^{[2]2}+P_Rg_i^2}\right)^2}{\sum_{i=1}^R\frac{g_i^2f_i^{[1]2}}{1+P_1f_i^{[1]2}+P_2f_i^{[2]2}+P_Rg_i^2}}
\end{eqnarray}
Substituting into (\ref{eq:R1R2def}) we have the expression for $C^{10}_2$.\hfill\QED
\subsection{Proof of Theorem \ref{theorem:sumrate}}\label{app:sumrate}
For sum rate we have $\mu_1=\mu_2=1$, i.e. $\mu_1'=0, \mu_2=1$. Substituting these values into (\ref{eq:key})  and defining $\mbox{SNR}^\star = \mbox{SNR}_1^\star +\mbox{SNR}_2^\star$, we have
{\allowdisplaybreaks
\begin{eqnarray}
& &  d_k\frac{\mbox{SNR}^\star}{P_R} = \frac{P_1\left(\sum gdf^{[1]}\right)g_kf_k^{[1]}+P_2\left(\sum gdf^{[2]}\right)g_kf_k^{[2]}}{1+P_1f_k^{[1]2}+P_2f_k^{[2]2}+P_Rg_k^2}\label{eq:dksumrate}\\
&\Rightarrow & \left(\sum gdf^{[1]}\right)\frac{\mbox{SNR}^\star}{P_R}=P_1A_{11}\left(\sum gdf^{[1]}\right) + P_2 A_{12} \left(\sum gdf^{[2]}\right)\\
&\mbox{and} & \left(\sum gdf^{[2]}\right)\frac{\mbox{SNR}^\star}{P_R}=P_1A_{12}\left(\sum gdf^{[1]}\right) + P_2 A_{22} \left(\sum gdf^{[2]}\right)\\
&\Rightarrow &\frac{\mbox{SNR}^\star}{P_R}=P_1A_{11} + P_2 A_{12} \frac{\left(\sum gdf^{[2]}\right)}{\left(\sum gdf^{[1]}\right)}\label{eq:gdfsolve}\\
&\mbox{and}&\frac{\mbox{SNR}^\star}{P_R}=P_1A_{12} \frac{\left(\sum gdf^{[1]}\right)}{\left(\sum gdf^{[2]}\right)}+ P_2 A_{22} 
\end{eqnarray}
}
Eliminating the factor $\frac{\left(\sum gdf^{[1]}\right)}{\left(\sum gdf^{[2]}\right)}$ from the last two equations leads to the quadratic equation:
\begin{eqnarray}
&&\left(\frac{\mbox{SNR}^\star}{P_R}\right)^2 - \left(P_1A_{11}+P_2A_{22}\right)\frac{\mbox{SNR}^\star}{P_R}+P_1P_2\left(A_{11}A_{22}-A_{12}^2\right) = 0\\
&\Rightarrow& \frac{\mbox{SNR}^\star}{P_R} = \frac{P_1A_{11}+P_2A_{22}\pm\sqrt{\left(P_1A_{11}+P_2A_{22}\right)^2-4P_1P_2\left(A_{11}A_{22}-A_{12}^2\right)}}{2}
\end{eqnarray}
The smaller root of the quadratic equation represents the \emph{minimum} value of the objective function, while the larger root is the desired \emph{maximum} value, which gives us the sum rate capacity expression of Theorem \ref{theorem:sumrate}.

From definition, $\tan\theta = \frac{\left(\sum dgf^{[2]}\right)}{\left(\sum dgf^{[1]}\right)}$ and its value is computed from (\ref{eq:gdfsolve}). 

Finally, we obtain the sum-rate optimal rate pairs by finding out $\mbox{SNR}_1^\star$ and $\mbox{SNR}_2^\star$ as follows.
{\allowdisplaybreaks
\begin{eqnarray}
\mbox{SNR}^\star&=&\frac{P_1\left(\sum gdf^{[1]}\right)^2+P_2\left(\sum gdf^{[2]}\right)^2}{1+\sum d^2g^2}, \\
\mbox{SNR}^\star_1&=&\frac{P_1\left(\sum gdf^{[1]}\right)^2}{1+\sum d^2g^2},~~\mbox{SNR}^\star_2=\frac{P_2\left(\sum gdf^{[2]}\right)^2}{1+\sum d^2g^2}\\
\mbox{Define}~~~~\beta &=& \frac{\mbox{SNR}^\star_1}{\mbox{SNR}^\star}~~~\mbox{so}~~~\mbox{SNR}^\star_1=\beta \mbox{SNR}^\star, \mbox{SNR}^\star_2=(1-\beta)\mbox{SNR}^\star\\
\Rightarrow \beta &= & \frac{P_1\left(\sum gdf^{[1]}\right)^2}{P_1\left(\sum gdf^{[1]}\right)^2+P_2\left(\sum gdf^{[2]}\right)^2}=\frac{P_1\frac{\left(\sum gdf^{[1]}\right)}{\left(\sum gdf^{[2]}\right)}}{P_1\frac{\left(\sum gdf^{[1]}\right)}{\left(\sum gdf^{[2]}\right)}+P_2\frac{\left(\sum gdf^{[2]}\right)}{\left(\sum gdf^{[1]}\right)}}
\end{eqnarray}
}
Substituting the value of $\frac{\left(\sum gdf^{[1]}\right)}{\left(\sum gdf^{[2]}\right)}$ from (\ref{eq:gdfsolve}) we have the definition of $\beta$ as stated in Theorem \ref{theorem:sumrate}.\hfill\QED
\subsection{Proof of Theorem \ref{theorem:mu1mu2}}\label{app:mu1mu2}
Continuing from (\ref{eq:key}),
\begin{eqnarray}
&& d_i \frac{\mu_2(1+\mbox{SNR}_1^\star)(\mbox{SNR}_1^\star+\mbox{SNR}_2^\star)+\mu_1'(1+\mbox{SNR}_1^\star+\mbox{SNR}_2^\star)(\mbox{SNR}_1^\star)}{P_R}=\nonumber\\
&&[\mu_1'(1+\mbox{SNR}_1^\star+\mbox{SNR}_2^\star)+\mu_2(1+\mbox{SNR}_1^\star)]P_1(\sum_{k=1}^Rd_kg_kf_k^{[1]})\frac{g_if_i^{[1]}}{1+P_1f_i^{[1]2}+P_2f_i^{[2]2}+P_Rg_i^2}\nonumber\\
&&+\mu_2(1+\mbox{SNR}_1^\star)P_2(\sum_{k=1}^Rd_kg_kf_k^{[2]})\frac{g_if_i^{[2]}}{1+P_1f_i^{[1]2}+P_2f_i^{[2]2}+P_Rg_i^2} 
\end{eqnarray}
Multiplying both sides of the equation with $g_if_i^{[1]}$ and summing up all the equations for $i=1$ to $i=R$, we have:
\begin{eqnarray}
&&\mu_2(1+\mbox{SNR}_1^\star)(\mbox{SNR}_1^\star+\mbox{SNR}_2^\star)+\mu_1'(1+\mbox{SNR}_1^\star+\mbox{SNR}_2^\star)(\mbox{SNR}_1^\star)=\nonumber\\
&&~~~~~~~~~~~~~~~~~~~~~~~\mu_2(1+\mbox{SNR}_1^\star)(P_RP_1A_{11}+P_RP_2A_{12}/\alpha)+\mu_1'(1+\mbox{SNR}_1^\star+\mbox{SNR}_2^\star)P_RP_1A_{11}~~~~~~~~~
\end{eqnarray}
where $\alpha=\frac{\left(\sum dgf^{[1]}\right)}{\left(\sum dgf^{[2]}\right)}=\sqrt{\frac{P_1\mbox{SNR}_1^\star}{P_2\mbox{SNR}_2^\star}}$.
Similarly,
\begin{eqnarray}
&&\mu_2(1+\mbox{SNR}_1^\star)(\mbox{SNR}_1^\star+\mbox{SNR}_2^\star)+\mu_1'(1+\mbox{SNR}_1^\star+\mbox{SNR}_2^\star)(\mbox{SNR}_1^\star)=\nonumber\\
&&~~~~~~~~~~~~~~~~~~~~~~~\mu_2(1+\mbox{SNR}_1^\star)(P_RP_1A_{12}\alpha+P_RP_2A_{22})+\mu_1'(1+\mbox{SNR}_1^\star+\mbox{SNR}_2^\star)P_RP_1A_{12}\alpha~~~~~~~~~
\end{eqnarray}
Simple algebraic manipulation of these two simultaneous equations gives us the result of Theorem \ref{theorem:mu1mu2}.\hfill\QED
\subsection{Proof of Theorems \ref{theorem:macbc} and \ref{theorem:macbc2}}\label{app:macbc}
Starting with the normalized BC as in (\ref{eq:bcnormal}), without loss of generality we assume
\begin{eqnarray}
\frac{[\mbox{Tr}\left({\bf F}^{[1]}{\bf D}{\bf G}\right)]^2{P}}{{\mbox{Tr}\left[{\bf D}^2\left(I+P{\bf F}^{[1]2}+P_R{\bf G}^2\right)\right]}}\geq  \frac{[\mbox{Tr}\left({\bf F}^{[2]}{\bf D}{\bf G}\right)]^2{P}}{{\mbox{Tr}\left[{\bf D}^2\left(I+P{\bf F}^{[2]2}+P_R{\bf G}^2\right)\right]}}
\end{eqnarray}
Thus, user $1$ is the stronger user in this degraded AWGN broadcast channel.

We wish to show that any rate pair $(R_1,R_2)$ achievable in MAC$({\bf F}^{[1]}, P_1, {\bf F}^{[2]}, P_2, {\bf G}, {\bf D}, P_R)$ is also achievable in BC$\left({\bf G}, P_R, {\bf F}^{[1]}, {\bf F}^{[2]}, \kappa{\bf D}, P_1+P_2\right)$. For this we need the following lemma.
\begin{lemma} \label{lemma1} Consider the following rate pair $(R_1, R_2)$ on the boundary of the capacity region of MAC$({\bf F}^{[1]}, P_1, {\bf F}^{[2]}, P_2, {\bf G}, {\bf D}, P_R)$ with
\begin{eqnarray}
R_1^{MAC}&=&\log\left(1+\frac{P_1P_R[\mbox{Tr}\left({\bf G}{\bf D}{\bf F}^{[1]}\right)]^2}{{\mbox{Tr}\left[{\bf D}^2\left(I+P_1{\bf F}^{[1]2}+P_2{\bf F}^{[2]2}+P_R{\bf G}^2\right)\right]}+P_2P_R[\mbox{Tr}\left({\bf G}{\bf D}{\bf F}^{[2]}\right)]^2}\right)\\
R_2^{MAC}&=&\log\left(1+\frac{P_2P_R[\mbox{Tr}\left({\bf G}{\bf D}{\bf F}^{[2]}\right)]^2}{{\mbox{Tr}\left[{\bf D}^2\left(I+P_1{\bf F}^{[1]2}+P_2{\bf F}^{[2]2}+P_R{\bf G}^2\right)\right]}}\right)
\end{eqnarray}
If this rate pair is achievable by \emph{any} scalar AWGN broadcast channel where user 1 is stronger than user 2, then \emph{every} achievable rate pair of MAC$({\bf F}^{[1]}, P_1, {\bf F}^{[2]}, P_2, {\bf G}, {\bf D}, P_R)$ is also achievable in that broadcast channel.
\end{lemma}
\begin{figure}[h]
\input{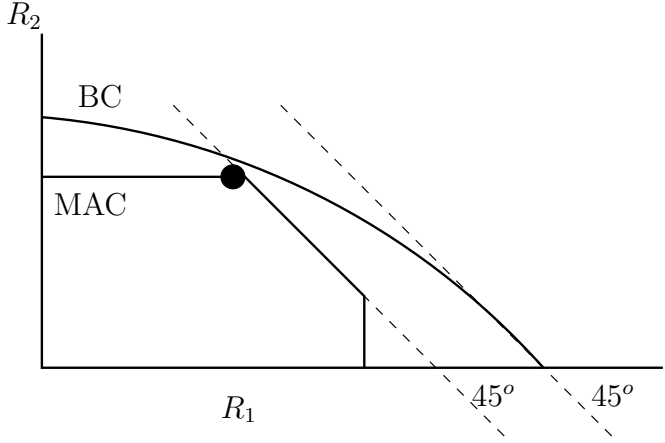}
\caption{Proof of Theorem \ref{lemma1}}\label{fig:lemmaproof}
\end{figure}
\begin{proof}The proof of Lemma \ref{lemma1} is evident from Fig. \ref{fig:lemmaproof}. The $45^o$ slope lines are tangent on a capacity region only at the sum rate maximizing point. For the MAC this is the slanted edge of the pentagon, while for the BC it is the point where all the rate and power is allocated to the stronger user, i.e. user $1$ in the figure. As is evident from the figure, if the marked corner of the MAC capacity region is inside the capacity region of the BC then the entire MAC capacity region must be contained within the capacity region of the BC.\hfill \QED
\end{proof}

Therefore, we only need to show that the rate pair $(R_1, R_2)$ specified in Theorem \ref{lemma1} is achievable in BC$\left({\bf G}, P_R, {\bf F}^{[1]}, {\bf F}^{[2]}, \kappa{\bf D}, P_1+P_2\right)$. Now, from the capacity of the scalar AWGN BC, we know every rate pair on the boundary of the capacity region of BC$\left({\bf G}, P_R, {\bf F}^{[1]}, {\bf F}^{[2]}, \kappa{\bf D}, P_1+P_2\right)$ can be written as:
\begin{eqnarray}
R_1^{BC}&=&\log\left(1+\frac{\alpha P_RP[\mbox{Tr}\left({\bf F}^{[1]}{\bf D}{\bf G}\right)]^2}{{\mbox{Tr}\left[{\bf D}^2\left(I+P{\bf F}^{[1]2}+P_R{\bf G}^2\right)\right]}}\right)\\
R_2^{BC}&=&\log\left(1+\frac{(1-\alpha) P_RP[\mbox{Tr}\left({\bf F}^{[2]}{\bf D}{\bf G}\right)]^2}{\alpha P_RP[\mbox{Tr}\left({\bf F}^{[2]}{\bf D}{\bf G}\right)]^2+{\mbox{Tr}\left[{\bf D}^2\left(I+P{\bf F}^{[2]2}+P_R{\bf G}^2\right)\right]}}\right)
\end{eqnarray}
Now, setting $R_1^{MAC}=R_1^{BC}$ we obtain the value of $\alpha$
\begin{eqnarray}
\alpha&=&\frac{P_1\mbox{Tr}\left[{\bf D}^2\left(I+P{\bf F}^{[1]2}+P_R{\bf G}^2\right)\right]}{P\mbox{Tr}\left[{\bf D}^2\left(I+P_1{\bf F}^{[1]2}+P_2{\bf F}^{[2]2}+P_R{\bf G}^2\right)\right]+PP_2P_R[\mbox{Tr}\left({\bf F}^{[2]}{\bf D}{\bf G}\right)]^2}
\end{eqnarray}
It is easily verified that $0\leq\alpha\leq 1$ so it is a feasible power split for the broadcast channel. 

Similarly, setting $R_2^{MAC}=R_2^{BC}$ we again solve for $\alpha$ to obtain:
\begin{eqnarray}
&&(1-\alpha)=\frac{\alpha P_RPP_2[\mbox{Tr}\left({\bf F}^{[2]}{\bf D}{\bf G}\right)]^2+P_2{\mbox{Tr}\left[{\bf D}^2\left(I+P{\bf F}^{[2]2}+P_R{\bf G}^2\right)\right]}}{P\mbox{Tr}\left[{\bf D}^2\left(I+P_1{\bf F}^{[1]2}+P_2{\bf F}^{[2]2}+P_R{\bf G}^2\right)\right]}\\
&\Rightarrow& \alpha\left(PP_RP_2 \left[\mbox{Tr}\left({\bf F}^{[2]}{\bf D}{\bf G}\right)\right]^2 + P\mbox{Tr}\left[{\bf D}^2\left(I+P_1{\bf F}^{[1]2}+P_2{\bf F}^{[2]2}+P_R{\bf G}^2\right)\right]\right)\nonumber\\
&&=P\mbox{Tr}\left[{\bf D}^2\left(I+P_1{\bf F}^{[1]2}+P_2{\bf F}^{[2]2}+P_R{\bf G}^2\right)\right]- P_2{\mbox{Tr}\left[{\bf D}^2\left(I+P{\bf F}^{[2]2}+P_R{\bf G}^2\right)\right]}~~~\\
\alpha&=&\frac{P_1\mbox{Tr}\left[{\bf D}^2\left(I+P{\bf F}^{[1]2}+P_R{\bf G}^2\right)\right]}{P\mbox{Tr}\left[{\bf D}^2\left(I+P_1{\bf F}^{[1]2}+P_2{\bf F}^{[2]2}+P_R{\bf G}^2\right)\right]+PP_2P_R[\mbox{Tr}\left({\bf F}^{[2]}{\bf D}{\bf G}\right)]^2}
\end{eqnarray}
Thus, the same value of $\alpha$ satisfies both $R_1^{MAC}=R_1^{BC}$ and $R_2^{MAC}=R_2^{BC}$. Therefore, with this value of $\alpha$, BC$\left({\bf G}, P_R, {\bf F}^{[1]}, {\bf F}^{[2]}, \kappa{\bf D}, P_1+P_2\right)$ can achieve the rate pair $(R_1^{MAC}, R_2^{MAC})$ specified in Theorem \ref{lemma1}. Applying the result of Theorem \ref{lemma1}, this implies that the capacity region of MAC$({\bf F}^{[1]}, P_1, {\bf F}^{[2]}, P_2, {\bf G}, {\bf D}, P_R)$ is contained within the capacity region of its dual BC. This establishes Theorem \ref{theorem:macbc}.

Theorem \ref{theorem:macbc2} is also easily established from the above result. This is because we have shown not only that the dual BC achieves the chosen rate pair on the boundary of the MAC, but also that this rate pair is on the boundary of the capacity region of the dual BC. In other words, we have shown above, that any choice of $P_1,P_2$ on MAC$({\bf F}^{[1]}, P_1, {\bf F}^{[2]}, P_2, {\bf G}, {\bf D}, P_R)$ achieves a point \emph{on the boundary} of the capacity region of BC$\left({\bf G}, P_R, {\bf F}^{[1]}, {\bf F}^{[2]}, \kappa{\bf D}, P_1+P_2\right)$. Moreover $P_1, P_2$ are related to $\alpha$ by a continuous function. Now, as we take $P_1$ from $0$ to $P$ while keeping $P_1+P_2=P$, we trace the entire boundary of the capacity region of BC$\left({\bf G}, P_R, {\bf F}^{[1]}, {\bf F}^{[2]}, \kappa{\bf D}, P_1+P_2\right)$. This implies the result of Theorem \ref{theorem:macbc2}.\hfill\QED

\section{Proof of Theorem \ref{theorem:macbcgeneral}}\label{app:macbcgeneral}
We begin by incorporating the power constraints into the channel definition. For the MAC, including the power constraint for the relays with total power $P_{R2}$ into the channel we have:
\begin{eqnarray}
y= \frac{\sqrt{P_{R2}}\overline{\bf G}^T{\bf B}_m{\bf H}{\bf A}_m\overline{\bf F}^{[1]}x_1+\sqrt{P_{R2}}\overline{\bf G}^T{\bf B}_m{\bf H}{\bf A}_m\overline{\bf F}^{[2]}x_2}{\sqrt{\scriptstyle{P_1 ||{\bf B}_m{\bf H}{\bf A}_m \overline{\bf F}^{[1]}||^2+P_2 ||{\bf B}_m{\bf H}{\bf A}_m \overline{\bf F}^{[2]}||^2+P_{R2}||\overline{\bf G}^T{\bf B}_m{\bf H}{\bf A}_m||^2+||{\bf B}_m{\bf H}{\bf A}_m ||^2+P_{R2}||\overline{\bf G}^T{\bf B}_m||^2+||{\bf B}_m||^2}}}~~+n \nonumber
\end{eqnarray}
Next, incorporating the power constraint for the relays with total power $P_{R1}$ into the channel we have:
\begin{eqnarray}
y= \sqrt{\frac{P_{R1}P_{R2}}{\Delta_m}}\overline{\bf G}^T{\bf B}_m{\bf H}{\bf A}_m\overline{\bf F}^{[1]}x_1+\sqrt{\frac{P_{R1}P_{R2}}{\Delta_m}}\overline{\bf G}^T{\bf B}_m{\bf H}{\bf A}_m\overline{\bf F}^{[2]}x_2~+n 
\end{eqnarray}
where
\begin{eqnarray}
\Delta_m&=&{P_1 P_{R1}||{\bf B}_m{\bf H}{\bf A}_m \overline{\bf F}^{[1]}||^2+P_2 P_{R1} ||{\bf B}_m{\bf H}{\bf A}_m \overline{\bf F}^{[2]}||^2+P_{R2}P_{R1}||\overline{\bf G}^T{\bf B}_m{\bf H}{\bf A}_m||^2+P_{R1}||{\bf B}_m{\bf H}{\bf A}_m ||^2}\nonumber\\
&&+P_1P_{R2}||\overline{\bf G}^T{\bf B}_m||^2||{\bf A}_m\overline{\bf F}^{[1]}||^2 +P_2P_{R2}||\overline{\bf G}^T{\bf B}_m||^2||{\bf A}_m\overline{\bf F}^{[2]}||^2 + P_{R2}||\overline{\bf G}^T{\bf B}_m||^2||{\bf A}_m||^2\nonumber\\
&&+P_1||{\bf B}_m||^2||{\bf A}_m\overline{\bf F}^{[1]}||^2 +P_2||{\bf B}_m||^2||{\bf A}_m\overline{\bf F}^{[2]}||^2 + ||{\bf B}_m||^2||{\bf A}_m||^2\nonumber
\end{eqnarray}

Similarly, incorporating the power constraint for the relays with power $P$ into the broadcast channel we get:
\begin{eqnarray}
y_1 &=&\frac{(\sqrt{P})\overline{\bf F}^{[1]T}{\bf A}_b{\bf H}^T{\bf B}_b\overline{\bf G}}{\sqrt{P_{R2}||{\bf A}_b{\bf H}^T{\bf B}_b\overline{\bf G}||^2 +||{\bf A}_b{\bf H}^T{\bf B}_b||^2+||{\bf A}_b||^2+P||\overline{\bf F}^{[1]T}{\bf A}_b||^2+P||\overline{\bf F}^{[1]T}{\bf A}_b{\bf H}^T{\bf B}_b||^2}}~~ x+n_1,\nonumber
\end{eqnarray}
and incorporating the power constraint for the relays with power $P_{R1}$ into the channel we get:
\begin{eqnarray}
y_1 &=&\sqrt{\frac{{PP_{R1}}}{\Delta_b^{[1]}}}\overline{\bf F}^{[1]T}{\bf A}_b{\bf H}^T{\bf B}_b\overline{\bf G}~ x+n_1,\nonumber
\end{eqnarray}
where
\begin{eqnarray}
\Delta_b^{[1]}&=&P_{R1}P_{R2}||{\bf A}_b{\bf H}^T{\bf B}_b\overline{\bf G}||^2 +P_{R1}||{\bf A}_b{\bf H}^T{\bf B}_b||^2+PP_{R1}||\overline{\bf F}^{[1]T}{\bf A}_b{\bf H}^T{\bf B}_b||^2\nonumber\\
&&+PP_{R2}||\overline{\bf F}^{[1]T}{\bf A}_b||^2||{\bf B}_b\overline{\bf G}||^2+P||\overline{\bf F}^{[1]T}{\bf A}_b||^2||{\bf B}_b||^2+P_{R2}||{\bf A}_b||^2||{\bf B}_b\overline{\bf G}||^2+||{\bf A}_b||^2||{\bf B}_b||^2\nonumber
\end{eqnarray}
and $y_2$ and $\Delta_b^{[2]}$ are similarly defined.

Proceeding as in the proof of Theorem \ref{theorem:macbc}, let us assume, without loss of generality that user $1$ is the stronger user on the resulting BC., i.e.,
\begin{eqnarray}
\frac{{PP_{R1}}}{\Delta_b^{[1]}}||\overline{\bf F}^{[1]T}{\bf A}_b{\bf H}^T{\bf B}_b\overline{\bf G} ||^2 \geq \frac{{PP_{R1}}}{\Delta_b^{[2]}}||\overline{\bf F}^{[2]T}{\bf A}_b{\bf H}^T{\bf B}_b\overline{\bf G}||^2
\end{eqnarray}
According to Lemma \ref{lemma1} we need to show that the following rate pair
\begin{eqnarray}
R_1^{MAC}&=&\log\left(1+\frac{P_1P_{R1}P_{R2} ||\overline{\bf G}^T{\bf B}_m{\bf H}{\bf A}_m\overline{\bf F}^{[1]} ||^2      }{\Delta_m+P_2P_{R1}P_{R2} ||\overline{\bf G}^T{\bf B}_m{\bf H}{\bf A}_m\overline{\bf F}^{[2]} ||^2 }\right)\\
R_2^{MAC}&=&\log\left(1+\frac{P_2P_{R1}P_{R2} ||\overline{\bf G}^T{\bf B}_m{\bf H}{\bf A}_m\overline{\bf F}^{[2]} ||^2      }{\Delta_m }\right)\\
\end{eqnarray}
is achievable and is on the boundary of the dual BC.

Rate pairs on the boundary of the degraded dual BC with user $1$ stronger than user $2$ are given by:
\begin{eqnarray}
R_1^{BC}&=&\log\left(1+{\alpha\frac{{PP_{R1}P_{R2}}}{\Delta_b^{[1]}}}||\overline{\bf F}^{[1]T}{\bf A}_b{\bf H}^T{\bf B}_b\overline{\bf G}||^2~ \right)\\
R_2^{BC}&=&\log\left(1+\frac{(1-\alpha)PP_{R1}P_{R2}||\overline{\bf F}^{[2]T}{\bf A}_b{\bf H}^T{\bf B}_b\overline{\bf G}||^2~}{\Delta_b^{[2]}+\alpha PP_{R1}P_{R2} ||\overline{\bf F}^{[2]T}{\bf A}_b{\bf H}^T{\bf B}_b\overline{\bf G}||^2  } \right)
\end{eqnarray}
\begin{eqnarray}
\alpha & = & \frac{\Delta_b^{[1]}P_1}{P\Delta_m+PP_2P_{R1}P_{R2} ||\overline{\bf G}^T{\bf B}_m{\bf H}{\bf A}_m\overline{\bf F}^{[2]} ||^2 }
\end{eqnarray}
\begin{eqnarray}
\alpha&=& \frac{P\Delta_m - P_2\Delta_b^{[2]}}{{P\Delta_m }+ PP_2P_{R1}P_{R2} ||\overline{\bf F}^{[2]T}{\bf A}_b{\bf H}^T{\bf B}_b\overline{\bf G}||^2}
\end{eqnarray}
It is easily verified that $P\Delta_m = P_1\Delta_b^{[1]}+P_2\Delta_b^{[2]}$. Thus, the same value of $\alpha$ satisfies both $R_1^{MAC}=R_1^{BC}$ and $R_2^{MAC}=R_2^{BC}$ and the chosen boundary point of the MAC is also shown to be a boundary point of the dual BC. By the same arguments as the proof of Theorem \ref{theorem:macbc}, the duality relationship is established. \hfill\QED
\bibliographystyle{ieeetr}
\bibliography{Thesis}
\end{document}